\documentclass[a4paper,11pt]{article}
\usepackage{pos}
\usepackage{epsfig}

\title{$SO(10)$-inspired leptogenesis}

\author*[a]{Pasquale Di Bari}

\affiliation[a]{School of Physics and Astronomy, University of Southampon\\
 Southampton, SO17 1BJ, U.K.}

\emailAdd{P.Di-Bari@soton.ac.uk}

\abstract{In the first part of the talk, I review general properties of $SO(10)$-inspired leptogenesis. This high-scale leptogenesis scenario is based 
on the simple assumption that the neutrino Dirac mass matrix is not too different from the up quark mass matrix. 
After showing how this necessarily implies a production of the asymmetry from the next-to-lightest 
right handed neutrino decays, so-called $N_2$-leptogenesis, I discuss how this results into important 
testable constraints on low energy neutrino parameters. In particular inverted ordering is not viable if strict
$SO(10)$-inspired conditions are assumed. This is an important test in view of the expected results from the JUNO experiment.
I also discuss how a subset of the $SO(10)$-inspired leptogenesis solutions 
realises strong thermal leptogenesis, where the final asymmetry is independent of the initial conditions. In this case
a signal might be discovered by next generation $0\nu\beta\beta$ decay experiments. 
In the second part, I present some new results from \cite{DiBari:2025zlv}, where the impact of flavour coupling 
on $SO(10)$-inspired leptogenesis has been studied in detail.}

\FullConference{Proceedings of the Corfu Summer Institute 2025 "School and Workshops on Elementary Particle Physics and Gravity" (CORFU2025)\\
27 April - 28 September, 2025\\
Corfu, Greece\\
Slides can be found at: https://www.southampton.ac.uk/~pdb1d08/PDB4CORFU25.pdf}

\begin{document}

\def\mc#1{\mathcal#1}
\def\a{\alpha}
\def\b{\beta}
\def\c{\chi}
\def\d{\delta}
\def\e{\epsilon}
\def\f{\phi}
\def\g{\gamma}
\def\h{\eta}
\def\i{\iota}
\def\j{\psi}
\def\k{\kappa}
\def\la{\lambda}
\def\m{\mu}
\def\n{\nu}
\def\o{\omega}
\def\p{\pi}
\def\q{\theta}
\def\r{\rho}
\def\s{\sigma}
\def\t{\tau}
\def\u{\upsilon}
\def\x{\xi}
\def\z{\zeta}
\def\D{\Delta}
\def\F{\Phi}
\def\G{\Gamma}
\def\J{\Psi}
\def\L{\Lambda}
\def\O{\Omega}
\def\P{\Pi}
\def\Q{\Theta}
\def\S{\Sigma}
\def\U{\Upsilon}
\def\X{\Xi}

\def\ve{\varepsilon}
\def\vf{\varphi}
\def\vr{\varrho}
\def\vs{\varsigma}
\def\vq{\vartheta}

\newcommand{\vev}[1]{\langle #1 \rangle}
\def\dg{\dagger}                                     
\def\ddg{\ddagger}                                   
\def\wt#1{\widetilde{#1}}                    
\def\mt{\widetilde{m}_1}
\def\mti{\widetilde{m}_i}
\def\mtj{\widetilde{m}_j}
\def\rt{\widetilde{r}_1}
\def\mtt{\widetilde{m}_2}
\def\mttt{\widetilde{m}_3}
\def\rtt{\widetilde{r}_2}
\def\mb{\overline{m}}
\def\VEV#1{\left\langle #1\right\rangle}        
\def\be{\begin{equation}}
\def\ee{\end{equation}}
\def\ds{\displaystyle}
\def\ra{\rightarrow}

\def\bea{\begin{eqnarray}}
\def\eea{\end{eqnarray}}

\maketitle

\section{Introduction}

Cosmological observations and neutrino oscillation experiments provide strong arguments in favour of the existence of new physics,
likely in the form of an extension of the standard model (SM). Cosmologically, the dark matter conundrum and the solution of the matter-antimatter
asymmetry of the universe cannot be explained with known physics. The explanation of neutrino masses and mixing, 
implied by neutrino oscillation experiments, is also requiring beyond SM physics. In addition to these traditional motivations, recent cosmological tensions and anomalies might also require new physics.
In this talk I first review $SO(10)$-inspired leptogenesis and then discuss recent results found in \cite{DiBari:2025zlv}, 
in collaboration with Xubin Hu, on the {\em Impact of flavour coupling on SO(10)-inspired leptogenesis}.
This is quite a general  scenario of high scale leptogenesis 
that can be realised within realistic grandunified models able to address the problem of understanding 
fermion masses and mixing with a reduced number of parameters. At the same time it makes different
{\em pre}dictions on low energy neutrino parameters, providing a clear example of a 
testable high scale leptogenesis scenario. 

\section{Neutrino masses and mixing parameters}

Before discussing $SO(10)$-inspired leptogenesis, I briefly discuss the current experimental situation in the determination 
of low energy neutrino parameters.  Let us denote the three neutrino masses by $m_1, m_2, m_3$. The corresponding mass eigenstates
are defined in a way that in the limit where the leptonic mixing matrix tends to the identity, then $\nu_1 = \nu_e$, $\nu_2 = \nu_\mu$ and $\nu_3 = \nu_\tau$. 
Neutrino oscillation experiments measure two neutrino mass squared differences, identifying two neutrino mass scales:
the solar neutrino mass scale $m_{\rm  sol} \equiv \sqrt{m^2_2 - m^2_1} = (8.6 \pm 0.1)\,{\rm meV}$ and
the atmospheric neutrino mass scale  $m_{\rm  atm} \equiv \sqrt{m^2_{3'} - m^2_{1'}} = (50.0 \pm 0.3)\,{\rm meV}$,
where  $m_{3'} = m_3 > m_2 > m_1 = m_{1'}$ in the case of normal ordering (NO) and
 $m_{1'} = m_3 < m_1 < m_2 =m_{3'}$ in the case of inverted ordering (IO). 
 This implies a lower bound 
 on the sum of neutrino masses $\sum_i m_i \geq 58\,{\rm meV}$ (at $95 \%$ C.L.) that is saturated in the NO case. 
 Most recent global analyses results, favour NO over IO at more than $3\sigma$ when atmospheric neutrino data are included \cite{Esteban:2026phq}.
Successful $SO(10)$-inspired leptogenesis is not attained in the IO case when low energy neutrino data are taken into account 
(I will be back on this point). 
Therefore, this agreement with the current results from global analyses is quite encouraging and it will be interesting if
it will be strengthened, to a higher statistical significance, by expected results from the JUNO experiment \cite{JUNO:2025gmd}.
In the following I will only refer to the NO case and in this case the lightest neutrino mass is $m_1$.  
 
The absolute neutrino mass scale can be parameterised in terms of $m_1$. 
This is not currently measured by absolute neutrino mass scale experiments that, however, 
place important upper bounds. First of all, assuming the $\Lambda$CDM model, 
cosmological observations combining CMB and BAO observations place a stringent upper bound on the 
sum of neutrino masses  given by $\sum_i m_i \leq 0.11 \, {\rm eV}$ (95 $\%$ CL) \cite{Tristram:2023haj}.
This translates into an upper bound on $m_1$ given by 
\be\label{upperbm1}
m_1 \lesssim 30\,{\rm meV} \,  .
\ee 
This has to be compared with the upper bound placed by neutrinoless double beta ($0\nu\b\b$) experiments 
on the effective  $0\nu\b\b$ neutrino mass $m_{ee} \equiv |m_{\nu ee}|$. 
The most stringent one has been set by the KamLAND-Zen collaboration, that finds \cite{KamLAND-Zen:2024eml}
$m_1 \leq (84 \mbox{--} 353) \, {\rm meV}$ (90\%\, {\rm C.L.}).  Clearly the cosmological upper bound 
(assuming $\Lambda$CDM) is much more stringent.\footnote{It is even more stringent
than it looks like, considering that the cosmological upper bound is at $95\%$ C.L. and the
upper bound from $0\nu\b\b$  at $90\%$ C.L. .} However, a  future $0\nu\b\b$ positive signal with a measurement of  $m_{ee}$ 
would not just provide information on the absolute neutrino mass scale but also some partial information on the Majorana phases
and, even more importantly, would be a clear discovery of lepton number violation at tree level, a crucial ingredient for leptogenesis. 
Finally, from Tritium beta decay, the KATRIN experiment has recently placed the upper bound 
$ m_1 \lesssim 0.45\,{\rm eV} \;\;\;  (90\% \,{\rm C.L.})$ \cite{Katrin:2024tvg}.

Let us now consider neutrino mixing. Parameterising the PMNS lepton mixing matrix $U$ in terms of the usual three mixing angles $\theta_{ij}$, 
the Dirac phase $\d$ and the Majorana phases $\rho$ and $\s$, 
\be\label{PMNS}
U =  \left( \begin{array}{ccc}
c_{12}\,c_{13} & s_{12}\,c_{13} & s_{13}\,e^{-{\rm i}\,\d} \\
-s_{12}\,c_{23}-c_{12}\,s_{23}\,s_{13}\,e^{{\rm i}\,\d} &
c_{12}\,c_{23}-s_{12}\,s_{23}\,s_{13}\,e^{{\rm i}\,\d} & s_{23}\,c_{13} \\
s_{12}\,s_{23}-c_{12}\,c_{23}\,s_{13}\,e^{{\rm i}\,\d}
& -c_{12}\,s_{23}-s_{12}\,c_{23}\,s_{13}\,e^{{\rm i}\,\d}  &
c_{23}\,c_{13}
\end{array}\right)
\, {\rm diag}\left(e^{i\,\rho}, 1, e^{i\,\sigma}
\right)\,   ,
\ee
where $s_{ij}\equiv \sin\theta_{ij}$ and $c_{ij}\equiv \cos\theta_{ij}$, 
latest global analyses of neutrino oscillation experimental results, including atmospheric neutrino
data from Super-Kamiokande and IceCube collaborations,  find for the mixing angles and 
leptonic Dirac phase $\d$ the following best fit values, $1\s$ errors  and $3\s$ intervals \cite{nufit24}: 
\bea\label{expranges}
\theta_{13} & = &  8.56^{\circ}\pm 0.11^{\circ} \in [8.19^{\circ}, 8.89^{\circ}] \,  , \\ \nonumber
\theta_{12} & = &  {33.68^{\circ}}^{+0.73^\circ}_{-0.70^\circ} \in [31.63^{\circ}, 35.95^{\circ}]  \,  , \\ \nonumber
\theta_{23} & = &  {43.3^{\circ}}^{+1.0^{\circ}}_{-0.8^\circ} \in [41.3^{\circ}, 49.9^{\circ}]  \,  ,  \\ \nonumber
\d & = &  {-148^{\circ}} ^{+26^{\circ}}_{-41^{\circ}} \in  [-236^{\circ}, 4^{\circ}]  \, .
\eea 
As one can notice, there is a $3\s$ exclusion interval , $\d\ni [4^\circ, 134^{\circ}]$, 
for the Dirac phase that disfavours $\sin\d > 0$. 

\section{Seesaw mechanism}

In order to describe neutrino masses, one could minimally extend the SM adding  three RH neutrinos and a Yukawa interaction term 
as for charged leptons.  In a generic lepton flavour basis, the total Lagrangian would then become ${\cal L} = {\cal L}^{SM}+{\cal L}_Y^{\nu}$, where
\be
-{\cal L}_Y^{\nu} = \overline{L}\,h \, \nu_{R}\, \widetilde{\Phi} \,  + {\rm h.c.}  \, ,
\ee
with the dual Higgs field defined as $\widetilde{\Phi} \equiv {\rm i}\,\sigma_2\,\Phi^{\star}$.
After electroweak spontaneous symmetry breaking, a neutrino Dirac mass matrix is generated by the Higgs vev $v$, 
$m_{D} =v\,h$,  so that the neutrino Dirac mass term in the Lagrangian would be written simply as
\be\label{Diracnumass}
-{\cal L}_{{\rm Dirac \, mass}}^{\nu} = \overline{\nu_{L}}\, m_{D}\,\nu_{R} + {\rm h.c.} \,  .
\ee
Analogously to the charged lepton mass term, this can be diagonalised by means of a bi-unitary transformation
(mathematically, the singular value decomposition of $m_D$)
\be\label{biunitary}
m_{D}= V^{\dagger}_L\, D_{m_D} \, U_R \,  ,
\ee
where $D_{m_D} \equiv {\rm diag}(m_{D1},m_{D2},m_{D3})$. In this way, one would simply have that the 
measured neutrino masses would be given by the Dirac neutrino masses, $m_i =m_{Di}$, and 
the leptonic mixing matrix $U = V_L^\dagger$. 
However, this minimal extension of the SM would leave unanswered different important questions:
\begin{itemize}
\item Why neutrinos are much lighter than all other fermions?
\item Why the  mixing angles in $U$ are so much larger than those in the CKM matrix?
\item How can we explain the dark matter and matter-antimatter asymmetry of the universe?
\item Why not to add a Majorana mass term as well?  
\end{itemize}
These questions can be addressed if one abandons  lepton number and even $B-L$ conservation at the classical level
introducing, in addition to the Yukawa coupling term, also a bare (not generated by radiative corrections) right-right Majorana mass term $M$.
One can then have in addition to the Dirac mass term also a right-right Majorana mass term, obtaining the seesaw Lagrangian
\be\label{Y+M}
- {\cal L}_{Y+M}^{\nu}=\overline{L}\,h\,\nu_{R}\, \widetilde{\Phi} \,   +
{1\over 2}\,\overline{\nu_{R}^{\,c}}\,M\,\nu_{R} + {\rm h.c.} \,  ,
\ee
that we are writing in the flavour basis where both charged lepton mass term and Mjorana mass term are diagonal.
The number of RH neutrinos $N$ should be regarded as a free parameter within this general phenomenological framework.
There is no model independent theoretical argument that can determine it. 
In particular, since RH neutrinos are SM singlets and do not carry a gauge anomaly, 
anomaly cancellation requirement does not apply. 
The Yukawa coupling matrix $h$ is then in general a $(3,N)$ matrix while the (right-right) Majorana
mass matrix $M$ is a $(N,N)$ matrix and notice that, since RH neutrinos are SM singlets, 
$M$ can also be a singlet and the term is renormalizable.
The Yukawa interaction defines the quantum number of the RH  neutrinos.
Assigning $L=+1$ to all  $\nu_R$ fields, the Dirac mass term conserves the total
lepton number  though not the individual lepton numbers or flavours.
However, though RH neutrinos carry a global charge, they do not carry 
gauge quantum numbers (colour, weak isospin, hypercharge). This is why they
can also have a Majorana mass term. This now breaks $L$ by two units: therefore, the introduction of a Majorana
mass term leads to lepton number non conservation at the classical level, with processes potentially
observable even at low energy such as $0\nu\b\b$ decay. It is also important to notice, being interested in leptogenesis, 
that it also clearly violates $B-L$. 
After spontaneous symmetry breaking, the neutrino
mass term can be now written in a generic flavour basis (we omit primed Latin indices) as 
\be
-{\cal L}_m^{\nu}= \overline{\nu_L}\,m_D\, \nu_R + 
{1\over 2}\,\overline{\nu^{c}_R}\,M \, \nu_{R} + {\rm h.c.}  \,  .
\ee
Using $\overline{\nu_L}\,m_D\, \nu_R = \overline{\nu^{c}_R}\,m_D^T\,\nu_L^{c}$, this can also be recast as
\be
-{\cal L}^{\nu}_{\rm m}= {1\over 2}\,
\left[(\overline{\nu_L},\overline{\nu_R^{c}})
\left(
\begin{array}{cc}
                0  & m_D  \\
               m_D^T &  M    \\
\end{array}\right)
\left(
\begin{array}{c}
       \nu_L^{c}  \\
       \nu_R  \\
\end{array}\right)
\right] + {\rm h.c.} \,  .
\ee
The most interesting case is the  {\em seesaw limit}  $M\gg m_D$.
In this case the spectrum splits into two sets:  
\begin{itemize}
\item A light neutrino set with masses given by the  seesaw formula \cite{seesaw}
\be\label{seesaw}
{\rm diag}(m_1,m_2,m_3) = U^\dagger \, m_D\,\mbox{\large ${1\over M}$}\,m_D^T \,  U^\star \,  ,
\ee
where
\be
m_{\nu} = - m_D\,\mbox{\large ${1\over M}$}\,m_D^T  \,  ,
\ee
is the low energy neutrino mass matrix (in the flavour basis).
\item A heavy neutrino set with masses approximately coinciding with the eigenvalues of $M$.
\end{itemize}
One can consider, as an illustrative example,  a one generation toy model. In this case the seesaw formula
reduces simply to  $M = m_D^2 /m_{\nu}$. Assuming $m_D \sim M_{EW} \sim 100\,{\rm GeV}$
and using $m_{\nu}\sim m_{\rm sol} \sim 10\,{\rm meV}$, one obtains $M \sim 10^{15}\,{\rm GeV}$.
This can be regarded as an encouraging exercise since it suggests that the light neutrino mass scale 
would not be a new fundamental scale but the result of an algebraic product, 
deriving from a combination of the electroweak scale and a scale very close to the grand-unified scale.

Let us now consider a realistic three generation case. In this case the RH neutrino mass spectrum
cannot be easily determined since the neutrino Dirac mass matrix contains  fifteen
parameters while, on the other hand, there are only nine low energy neutrino parameters in $m_{\nu}$ and 
we do not even test all of them considering that $0\nu\b\b$ decay would give information just on one quantity, $m_{ee}=|m_{\nu 11}|$,
involving two Majorana phases. However, we can consider two limit cases that are quite interesting since they
correspond to two well motivated classes of models.  For definiteness, we can consider the most attractive case $N=3$. 

A first limit case is obtained assuming that all mixing is generated by a mismatch between Yukawa and charged lepton bases while the Majorana
mass matrix is diagonal in the Yukawa basis. This assumption corresponds to take $U_R = P^{(ijK)}$ where $P^{(ijK)}$ is one 
of the six permutation matrices.\footnote{They can be expressed as: 
$P^{(ijK)}_{\ell m} = \d_{\ell i}\,\d_{m1} + \d_{\ell j}\,\d_{m2} + \d_{\ell K}\,\d_{m3}$.} 
Plugging the biunitary parameterisation Eq.~(\ref{biunitary})  with $U_R = P^{(ijK)}$ into the seesaw formula (\ref{seesaw}), one immediately finds as a solution
$U = V^\dagger_L$ for the leptonic mixing matrix and $m_i = m^2_{Dj}/M_{I}$ for the neutrino masses. The simplest case is $P^{(ijK)} =I$ and in that
case one has $m_i = m^2_{Di}/M_I$. If one further assumes that  the Yukawa spectrum is degenerate $m_{D1} = m_{D2} = m_{D3} = \lambda$, then one has
$m_i = \lambda^2/M_i$. This set of assumption ($U_R = I$ and degenerate spectrum) comes from having $m^\dagger_D m_D = \lambda^2 I$, a 
maximally symmetric case emerging from symmetry flavour models  \cite{Altarelli:2010gt,King:2013eh,Bertuzzo:2009im,DiBari:2018fvo}.
In this case the RH neutrino mass spectrum is given by $M_i = \lambda^2/m_i$, it is then inverted compared to the light neutrino mass spectrum
and hierarchical but not that strongly hierarchical. 
For example, taking $\lambda = 100\,{\rm GeV}$, $m_1 \sim 10^{-4}\,{\rm eV}$, $m_2 \simeq m_{\rm sol} ~ 10\,{\rm meV}$
and $m_3 \simeq m_{\rm atm} \sim 50\,{\rm meV}$, one finds $M_1 \sim 2\times 10^14\,{\rm GeV}$, $M_2 \sim 10^{15}\,{\rm GeV}$ 
and $M_3 \sim 10^{17}\,{\rm GeV}$. 

A second limit case is obtained assuming that all mixing stems from the RH neutrino Majorana mass, corresponding to take $V_L = I$. 
In addition, assuming $m_{D3} \gg m_{D2} \gg m_{D1}$, one finds for the RH neutrino mass spectrum \cite{Akhmedov:2003dg,decrypting}  
\be\label{MI}
M_1    \simeq   {m^2_{D1} \over |m_{\nu 11}|} \, , \;\;
M_{2}  \simeq    {m^2_{D2} \over m_1 \, m_2 \, m_3 } \, {|m_{\nu 11}| \over |(m_{\nu}^{-1})_{33}|  } \,  ,  \;\;
M_{3}  \simeq   m^2_{D3}\,|(m_{\nu}^{-1})_{33}|   \,  .
\ee 
For example, imposing $SO(10)$-inspired conditions \cite{SO10inspired,Branco:2002kt}
\be\label{alphas}
m_{D1} = \a_1 \, m_{\rm up}  \,  , \;\; 
m_{D2} =\a_2 \, m_{\rm charm} \,  , \;\;
m_{D3} = \a_3 \,m_{\rm top}  \,  ,
\ee
with $\a_i ={\cal O}(1)$, one finds, for $m_1 \sim m_{\rm sol} \sim 10\,{\rm meV}$:
\be\label{SO10spectrum}
M_1 \sim 10^5\,{\rm GeV} \,  \;  M_2 \sim 10^{10}\,{\rm GeV} \,  , \;  M_3 \sim 10^{15} \, {\rm GeV} \,   ,
\ee
a much more hierarchical spectrum than in the previous case.\footnote{From the analytical expressions Eq.~(\ref{MI}), one can 
derive the following expressions for the RH neutrino masses in the hierarchical limit (for $m_1 \ll  m_{\rm sol}$) \cite{Branco:2002kt}:
\be
M_1 \simeq {m^2_{D1} \over m_{\rm sol}\,s^2_{12}} \,   , \;\;
M_2 \simeq {m^2_{D2} \over m_{\rm atm}\,s^2_{23}} \,  ,  \;\;
M_3 \simeq  {m^2_{D3} \over m_1 } \,s^2_{23} \, s^2_{12} \,  .
\ee}  

Of course one can have all kind of situations in between and also play with
the absolute neutrino Dirac mass scale getting a much lighter RH neutrino mass spectrum, for example at the TeV scale. 
The question is then how we can test the existence of these very  heavy seesaw neutrinos and in case 
determine their masses.

\section{Minimal scenario of leptogenesis}

From a combined analysis of CMB anisotropies and BAO, the {\em Planck} collaboration finds for the baryonic
contribution to the energy density parameter \cite{Tristram:2023haj}
\be\label{OB0}
\Omega_{B0} h^2 = 0.02229 \pm 0.00012 \,  . 
\ee
All attempts to detect primordial antimatter in our observable universe have failed so far. We can then write
\be
\eta_{B0} \equiv {n_{B0} -  n_{\bar{B} 0} \over n_{\g 0}} 
\simeq {\O_{B0}\,\varepsilon_{\rm c 0}\over m_p \, n_{\gamma 0}}
\simeq 273.5\,\O_{B 0}h^2\,10^{-10} \,  ,
\ee
 so that from Eq.~(\ref{OB0}) one finds: 
\be\label{etaB0}
\eta_{B 0}^{\rm obs} = (6.10 \pm 0.03)\times 10^{-10} \,  .
\ee
This can be used as a value expressing the matter-antimatter asymmetry of the universe that needs to be reproduced by a baryogenesis model.
The Sakharov conditions are all satisfied in the standard model if the reheat temperature $T_{\rm RH} > T_{\rm sph}^{\rm out} \simeq 132\,{\rm GeV}$ 
\cite{DOnofrio:2014rug} in a way that the rate of sphaleron processes is not Boltzmann suppressed. This  is sufficient to generate some 
matter-antimatter asymmetry. However, any attempt to reproduce the observed matter-antimatter asymmetry fails by many orders of magnitude,
since the amount of $C\!P$ violation and the departure from thermal equilibrium are not sufficient. For this reason the observed matter-antimatter asymmetry
is regarded as a strong motivation for the existence of new physics. 

The seesaw mechanism extension of the SM provides the right ingredients to build a successful model of baryogenesis via leptogenesis \cite{fy}.
The decays of the seesaw neutrinos into lepton and Higgs doublets can violate $C\!P$ and occur out-of-equilibrium in a sufficient way to generate
a $B-L$ asymmetry. This is initially injected in the form of a lepton asymmetry but if 
$T_{\rm RH} > T_{\rm sph}^{\rm out} \simeq 132\,{\rm GeV}$, then part of the lepton asymmetry is converted into a baryon asymmetry.
In a minimal scenario of leptogenesis (see \cite{Blanchet:2012bk} for a review) the asymmetry is produced thermally, so that
the RH neutrinos are produced from the thermal bath and they decay at temperatures $T \sim M_I \lesssim T_{\rm RH}$,
and assuming the minimal (type-I) seesaw mechanism we discussed.  In this way the injected lepton asymmetry 
will be partly converted into a baryon asymmetry by sphaleron processes while conserving $B-L$.  In this way
the predicted baryon-to-photon ratio at the present time from leptogenesis can be expressed as
\be\label{etaB0}
\eta_{B0}^{\rm lep} = {a_{\rm sph} N_{B-L}^{\rm fin} \over N_{\gamma 0}} \simeq 0.01 \, N_{B-L}^{\rm fin}  \,  ,
\ee
where $a_{\rm sph} \simeq 1/3$ is the fraction of $B-L$ asymmetry ending up into a baryon asymmetry.
The last numerical expression holds if one normalises abundances in a  portion of comoving volume containing
one seesaw neutrino in ultra-relativistic thermal equilibrium. If flavour effects are negligible, 
the final $B-L$ asymmetry can be expressed as
\be\label{BmLf}
N_{B-L}^{\rm fin} = \sum_I^{1,2,3} \ve_I\,\kappa^{\rm f}(K_I) \, ,
\ee
where we defined the total $C\!P$ asymmetries as 
\be
\ve_I \equiv - {\Gamma(N_I \ra {L}_I + \phi^\dagger) - \Gamma(N_I \ra \bar{L}_I + \phi) \over 
\Gamma(N_I \ra {L}_I + \phi^\dagger) + \Gamma(N_I \ra \bar{L}_I + \phi) } \,   .
\ee
and $\kappa^{\rm f}(K_I)$ is the final {\em efficiency factor} depending on the total decay parameters
\be
K_I \equiv {\widetilde{\G}_{D, I} \over H(T=M_I)} \,  ,
\ee
where the decay widths are given by
\be\label{decaywidth}
\widetilde{\G}_{D, I} \equiv \Gamma_{D, I}(T=0) = {M_I\,v^2 \over 8\,\pi}\,(m^\dagger\,m)_{II}  
\ee
and $ \Gamma_{D, I} = \Gamma(N_I \ra {L}_I + \phi^\dagger) + \Gamma(N_I \ra \bar{L}_I + \phi)$.

The condition of successful leptogenesis is equivalent to impose $\eta_{B0}^{\rm lep}=\eta_{B0}^{\rm obs}$.
In general, $\eta_{B0}^{\rm lep}$ will be a function of all seesaw parameters, given by the 15 parameters in 
the Dirac neutrino mass matrix $m_D$ plus the 3 seesaw neutrino masses $M_I$, in total 18 parameters. 
The experimental information from low energy neutrino experiments is clearly insufficient to over constraint the
seesaw parameter space. This is clear if one considers the orthogonal parameterisation of the neutrino Dirac mass matrix \cite{Casas:2001sr}: 
\be
m_D = U \, \sqrt{D_m} \, \Omega \, \sqrt{D_M} \,  ,
\ee   
where $\Omega$ is an orthogonal complex matrix that, therefore, can be parameterised in therms of 6 parameters,
for example 3 complex angles. These 6 parameters, together with the three masses $M_I$, 
allow to determine the 3 total decay rates $\Gamma_I$ and 3 total $C\! P$ asymmetries $\ve_I$. 

The condition of successful leptogenesis provides an additional experimental constraint but, in general, this is still
insufficient to over constraint the seesaw parameter space. For this reason, there is not a way to test model independently the
seesaw mechanism  relying just on low energy neutrino parameters, even when leptogenesis is included. 
One needs to add more theoretical and/or experimental constraints to reduce
the number of independent parameters. There are two popular strategies:
\begin{itemize}
\item {\em low-scale seesaw and leptogenesis}: in this case one searches for a direct discovery of RH neutrinos 
in lab neutrino experiments;
\item {\em high-scale seesaw leptogenesis}: it is more challenging to test but there are a few strategies able to reduce
the number of parameters in order to obtain testable predictions on low energy neutrino parameters. 
\end{itemize}
In both cases one can search for additional experimental probes that can complement leptogenesis and low energy neutrino parameters
such as dark matter (see for example \cite{Asaka:2005pn,Anisimov:2008gg}) and, remarkably, 
primordial gravitational waves \cite{Dror:2019syi,DiBari:2020bvn,DiBari:2021dri,Fu:2022lrn,DiBari:2023mwu}. 

A simple traditional example of high scale leptogenesis scenario where a reduction of the number of seesaw parameters leads to a prediction on some 
low energy neutrino parameter, specifically the absolute neutrino mass scale, is given by so-called {\em vanilla leptogenesis} \cite{bounds}.  
Such a reduction relies on the following set of assumptions and approximations:
\begin{itemize}
\item[(i)] The flavour composition of leptons and anti-leptons produced in the decays and inverse decays of RH neutrinos does not 
influence the value of the final asymmetry and is, therefore, neglected;
\item[(ii)] The RH neutrino mass spectrum is hierarchical and, more precisely, $M_2 \gtrsim 2\,M_1$;
\item[(iii)] The lightest RH neutrino decay parameter $K_1 \gg 1$;
\item[(iv)] There are no fine-tuned cancellations in the seesaw formula.
\end{itemize}
Form the first three assumptions, one can conclude that the 
final asymmetry is dominated by the lightest RH neutrino ($N_1$-leptogenesis). 
In this way the expression (\ref{BmLf}) simplifies into
\be
N_{B-L}^{\rm f} \simeq N_{B-L}^{{\rm f}(N_1)}(m_1,M_1,K_1) = \ve_1(M_1,m_1)\,\kappa^{\rm f}(K_1,m_1) \,  ,
\ee 
in a way that the dependence of the four parameters 
associated to the asymmetry production from the two heavier RH neutrinos cancels out. 
In addition the total $C\!P$ asymmetry does not depend on the leptonic mixing matrix.  Moreover,
using the assumption (iv), one arrives to the upper bound \cite{di}
\be
\ve_1 \lesssim \bar{\ve}(M_1) \equiv {3 \over 16\,\pi}\, {M_1 \, m_{\rm atm} \over v^2} \simeq 10^{-6}\,
\left({M_1 \over 10^{10}\,{\rm GeV}} \right)  \, {m_{\rm atm} \over m_1 + m_3} \,  .
\ee
In this way one has an upper bound on the $B-L$ asymmetry depending just on three parameters: $K_1$, $m_{1}$, $M_1$.
Imposing the successful leptogenesis condition, one finds two important bounds on neutrino masses:
\begin{itemize}
\item A lower bound on the RH neutrino masses $M_1 \gtrsim 10^{9}\,{\rm GeV}$ \cite{di,cmb}. 
This also translates into a lower bound on the reheat temperature
of the universe that is slightly more relaxed since the asymmetry is produced quite sharply around a temperature 
$T_B \sim M_1/z_B \sim M_1/(2$--$5)$ \cite{Buchmuller:2004nz}.
\item An upper bound on the absolute neutrino mass scale $m_1 \lesssim 0.12\,{\rm eV}$ \cite{Buchmuller:2002jk,Buchmuller:2003gz,Buchmuller:2004nz,bounds,Garbrecht:2024xfs} .
\end{itemize}
The upper bound on the absolute neutrino mass scale is interesting since it is a clear example of how even for high scale leptogenesis 
models one can get predictions  on low energy neutrino parameters imposing certain set of assumptions that reduce the number of parameters. 
In the case of vanilla leptogenesis the $N_1$ dominance makes in a way that the four parameters describing properties of the 
two heavier RH neutrinos cancel out.\footnote{Notice that this predicted upper bound is nowadays confirmed by the upper bound from 
cosmological observations Eq.~(\ref{upperbm1}). Of course that is not enough to say vanilla leptogenesis is proved but it can be regarded 
as an encouraging result.}

The lower bound on $M_1$ seems to rule out a $SO(10)$-inspired leptogenesis scenario based on the $SO(10)$-inspired 
conditions in Eq.~(\ref{alphas}), leading to the RH neutrino mass spectrum in Eq.~(\ref{SO10spectrum}).  Before showing 
how this conclusion changes when flavour effects are considered, let us first discuss an(other) interesting property of the
simple vanilla leptogenesis scenario.

{\em Strong thermal leptogenesis}.  The strong washout assumption, $K_1 \gg 1$, is quite strongly supported by neutrino oscillation experiments. 
This is because the decay parameter can be recast as
$K_1 = \widetilde{m}_1 / m_\star$, where $\widetilde{m}_1 \equiv v^2\,(h^\dagger\,h)_{11}$ is the
{\em effective neutrino mass} \cite{plumacher} and
\be
m_{\star} \equiv  {16\,\pi^{5/2} \, \sqrt{g^{SM}_{\star}}   \over 3 \sqrt{5}} \, { v^2 \over   M_{\rm P}}  \simeq 1.07 \, {\rm meV}
\ee
is the {\em equilibrium neutrino mass} \cite{orloff}. The seesaw formula favours typically 
$\widetilde{m}_1$ in the range $m_{\rm sol}$--$m_{\rm atm} \sim (10$--$50)\,{\rm meV}$ corresponding to
$K_1 \sim 10$--$50$.\footnote{In this range a simple and very good analytical fit for the final efficiency factor is given by: 
\be\label{kappaflep}
\kappa_1^{\rm f}(K_1) \simeq {0.5 \over K_1^{1.2}}\,  .
\ee}
For this reason having $K_1 \lesssim 1$ is a very special situation.\footnote{Recently it was shown that for 
randomly uniformly (in flavour space) generated  seesaw models respecting neutrino oscillation experimental results, 
the probability to have $K_1\lesssim 1 $ is less than $0.1\%$ \cite{DiBari:2018fvo}.} This range of values of $K_1$ corresponds to
efficiency factors $\k^{\rm f} \sim 10^{-3}$--$10^{-2}$ that still allows successful leptogenesis if $M_1$ is above the 
lower bound. On the other hand for such values of $K_1$ the efficiency factor is independent of the initial $N_1$ abundance
and, moreover, any initial pre-existing asymmetry $N^{\rm p,i}_{B-L}$ is very efficiently washed-out, since its final value is exponentially 
suppressed, explicitly: 
\be\label{preexisting}
N^{\rm p,f}_{B-L} = N^{\rm p,i}_{B-L} \, e^{-{3\pi \over 8}\, K_1} \,  . 
\ee

\section{$N_2$-leptogenesis}

The washout of an initial pre-existing asymmetry from lightest RH neutrino decays in the vanilla leptogenesis scenario also applies to the washout 
of the asymmetry produced by $N_2$-decays. In this way the final asymmetry is dominated just by the contribution from $N_1$-decays. 
More explicitly, within the assumptions of vanilla leptogenesis, one can write the contribution to $\eta_{B0}$ from $N_2$-decays \cite{geometry} as:
\be
\eta_{B0}^{\rm lep (N_2)} \simeq 0.01\, \ve_2 \, \kappa^{\rm f}(K_2)\, e^{-{3\pi\over 8}K_1} \ll \eta_{B0}^{\rm obs} \,  .
\ee
However, in the case of $SO(10)$-inspired models, one has $M_1 \ll 10^9\,{\rm GeV}$ and in this case
flavour effects need to be taken into account \cite{flavoureffects}. In particular, for such low values of $M_1$, a three-flavoured regime
is realised implying that the wash-out occurs independently for each flavour.  Consequently, the final value of the $B-L$ asymmetry 
produced by $N_2$-decays, taking into account flavour effects, has now to be calculated as the sum of three contributions \cite{vives,bounds}:
\be
N_{B-L}^{{\rm f} (N_2)} =   
\ve_{2e}\,\kappa(K_{2e} + K_{2\mu})\, e^{-{3\pi\over 8}\,K_{1 e}} +
\ve_{2\mu}\,\kappa(K_{2e} + K_{2\mu})\, e^{-{3\pi\over 8}\,K_{1 \mu}} +
\ve_{2\tau}\,\kappa(K_{2\tau})\,e^{-{3\pi\over 8}\,K_{1 \tau}} \,  .
\ee
The flavour decay parameters are such that $K_{1e}+K_{1\mu}+K_{1\tau} = K_1$. In this way a single flavour decay parameter
can be much smaller than $K_1$.  This implies that the domain of successful $N_2$-leptogenesis now greatly enlarges. This is clearly 
illustrated by the results of a random scan taking into account neutrino oscillation experimental data, finding that while
the probability $p(K_1 <1) \lesssim 0.1 \%$, 
the probability that only one of the $K_{1\a} < 1$ is $\sim 50\%$ \cite{DiBari:2018fvo}. 
In this way the asymmetry produced by $N_2$-decays can survive much more easily and 
reproduce the final asymmetry, realising so-called $N_2$-leptogenesis. 

An interesting aspect of $N_2$-leptogenesis is that, while in the case of $N_1$-leptogenesis a two RH neutrino seesaw model would be
sufficient to reproduce the asymmetry, in the case of $N_2$-leptogenesis the existence of the heaviest RH neutrino $N_3$ is necessary if 
$M_1 \lesssim 10^{11}\,{\rm GeV}$ \cite{Antusch:2011nz}.  This is in order for the $\ve_{2\alpha}$'s not to be negligible. 

Another important aspect of $N_2$-leptogenesis is that it is the only hierarchical scenario that can realise strong thermal
leptogenesis when flavour effects are taken into account \cite{Bertuzzo:2010et}. 
In this case the asymmetry must be tauon-flavour dominated. Moreover, there is a very stringent lower bound
on the absolute neutrino mass scale depending on the initial value of the pre-existing asymmetry. For example
for a pre-existing asymmetry $N^{\rm p}_{B-L} \sim 10^{-3}$, one has $m_1 \gtrsim 10\,{\rm meV}$ \cite{DiBari:2014eqa}.

Finally, we are ready to answer the interesting question whether successful $N_2$ leptogenesis can 
be realised within $SO(10)$-inspired models and whether
it can be realised even in a strong thermal version. 

\section{$SO(10)$-inspired leptogenesis}

We have seen that starting from the $SO(10)$-inspired conditions  in Eq.~(\ref{alphas}) with $V_L = I$, 
one finds the analytical expressions  Eq.~(\ref{MI}) for the RH neutrino masses.  One can also find
the following analytical expression for the RH neutrino mixing matrix \cite{Akhmedov:2003dg,decrypting}:
\be\label{URapp}
U_R \simeq \left( \begin{array}{ccc}
1 & -{m_{D1}\over m_{D2}} \,  {m^\star_{\nu e \mu }\over m^\star_{\nu ee}}  & 
{m_{D1}\over m_{D3}}\,
{ (m_{\n}^{-1})^{\star}_{e\t}\over (m_{\n}^{-1})^{\star}_{\t\t} }   \\
{m_{D1}\over m_{D2}} \,  {m_{\nu e \mu }\over m_{\nu ee}} & 1 & 
{m_{D2}\over m_{D3}}\, 
{(m_{\n}^{-1})_{\m\t}^{\star} \over (m_{\n}^{-1})_{\t\t}^{\star}}  \\
 {m_{D1}\over m_{D3}}\,{m_{\nu e\t }\over m_{\nu e e}}  & 
- {m_{D2}\over m_{D3}}\, 
 {(m_\nu^{-1})_{\m\t}\over (m_\nu^{-1})_{\t\t}} 
  & 1 
\end{array}\right) \, D_{\Phi} ,
\ee
where $D_{\Phi} = {\rm diag}(\Phi_1,\Phi_2,\Phi_3)$ with
\be\label{RHphases}
\Phi_1 = {\rm Arg}[-m_{\nu ee}^{\star}] \, , \;  
\Phi_2 =  {\rm Arg}\left[{m_{\nu ee}\over (m_{\nu}^{-1})_{\t\t}}\right] -2\,(\rho+\s)  \,  , \;
\Phi_3 = {\rm Arg}[-(m_{\nu}^{-1})_{\t\t}] \,   .
\ee
In this way one trades off the unknown parameters in the $U_R$, with the three Dirac neutrino masses
and the nine low energy neutrino parameters in $m_\nu$. From these analytical expressions for the three $M_I$ and $U_R$,
one can also find analytical expressions for the flavoured decay parameters and the 
three $N_2$ flavoured $C\!P$ asymmetries, finding the following strong hierarchy:
\be\label{CPhierarchy}
\ve_{2\t}:\ve_{2\m}:\ve_{2e} = \a_3^{\,2}\,m^2_t : \a_2^{\, 2}\,m^2_c :
\a_1^{\,2}\,m^2_u \, {\a_3 m_t \over a_2 \,m_c} \, 
{\a_1^{\,2}\,m_u^2 \over \a_2^{\,2}\,m^2_c} \,  .
\ee 
This shows that the tauon $C\!P$ asymmetry is by far the dominant one and, ultimately, successful leptogenesis
can be achieved only when the asymmetry is tauon-dominated. In this way the final $B-L$ asymmetry is
dominated by the tauon flavour and can be written as
\bea
N_{B-L}^{\rm f} & \simeq & N_{\D_\tau}^{\rm f} \simeq  \ve_{2\tau}\,\kappa(K_{2\tau})\,e^{-{3\pi\over 8}\,K_{1 \tau}} \\
& \simeq & 
{3\over 16\,\pi}\, {\a_2^2\,m_c^2 \over v^2}\, {|m_{\nu ee}|\,
(|m^{-1}_{\nu \t \t}|^2 + |m^{-1}_{\nu \m \t}|^2)^{-1} \over m_1\,m_2\,m_3}\,
{|m^{-1}_{\n\t\t}|^2\over |m^{-1}_{\n\m\t}|^2}\,\sin\a_L    \\  \nonumber
& \times & \kappa\left({m_1\,m_2\,m_3 \over m_{\star}}\, 
{|(m_{\nu}^{-1})_{\m \t}|^2 \over |m_{\nu ee}|\, |(m_{\nu}^{-1})_{\t \t}|} \right)  \\  \nonumber
& \times & 
e^{-{3\pi\over 8}\,{|m_{\nu e\t}|^2 \over m_{\star}\,|m_{\nu ee}|}  }  \,  ,
\eea
where 
\be
\a_L =  {\rm Arg}\left[m_{\nu ee}\right]  - 2\,{\rm Arg}[(m^{-1}_{\nu})_{\m\t}] + \pi -2\,(\rho+\s)  
\ee 
can be regarded as the effective $SO(10)$-inspired leptogenesis phase.
From Eq.~(\ref{etaB0}) one can then obtain $\eta_{B0}^{\rm lep}$ as a function of the nine low energy neutrino parameters
and $\alpha_2$.  If one fixes $\alpha_2$ to some plausible maximum value, then 
the successful leptogenesis condition $\eta_{B0}^{\rm lep}(m_\nu, \alpha_2) = \eta_{B0}$ identifies 
an hypersurface in the space of the low energy neutrino parameters in a way that one obtains many different
kind of predictive conditions and connections among the various parameters.  What is of course non-trivial
is that current data do satisfy these conditions, so that successful $SO(10)$-inspired leptogenesis is
indeed compatible with the current experimental information on low energy neutrino parameters \cite{riotto1}.
The most remarkable condition on low energy neutrino parameters is the existence of a lower bound 
$m_1 \gtrsim 1 {\rm meV}$ on the absolute neutrino mass scale. 

Before giving more details on the different predictive constraints on low energy neutrino parameters,
let us generalise the discussion relaxing the simple condition $V_L = I$. We can parameterise
the unitary matrix $V_L$ using an analogous parameterisation as for the leptonic mixing matrix:
 \begin{equation}\label{Umatrix}
V_L=
\left( \begin{array}{ccc}
c^L_{12}\,c^L_{13} & s^L_{12}\,c^L_{13} & s^L_{13}\,e^{-{\rm i}\,\d_L} \\
-s^L_{12}\,c^L_{23}-c^L_{12}\,s^L_{23}\,s^L_{13}\,e^{{\rm i}\,\d_L} &
c^L_{12}\,c^L_{23}-s^L_{12}\,s^L_{23}\,s^L_{13}\,e^{{\rm i}\,\d_L} & s^L_{23}\,c^L_{13} \\
s^L_{12}\,s^L_{23}-c^L_{12}\,c^L_{23}\,s^L_{13}\,e^{{\rm i}\,\d_L}
& -c^L_{12}\,s^L_{23}-s^L_{12}\,c^L_{23}\,s^L_{13}\,e^{{\rm i}\,\d_L}  &
c^L_{23}\,c^L_{13}
\end{array}\right)
\, {\rm diag}\left(e^{i\,\rho_L}, 1, e^{i\,\sigma_L}  \right)\, ,
\end{equation}
where $s^L_{ij} \equiv \sin\theta_{ij}^L$ and $c^L_{ij}\equiv \cos\theta_{ij}^L$. By definition of 
$SO(10)$-inspired leptogenesis, we can impose $0 \leq \theta_{ij}^L \lesssim \theta_{ij}^{CKM}$,
where $\theta_{ij}^{CKM}$ are the quark mixing angles in the CKM matrix. 
The results will not depend on variations of the upper bounds on the angles in the $V_L$ matrix by some ${\cal O}(1)$ factors . 

An analytical description can be still obtained also in this more general case \cite{full}. Introducing
the light neutrino mass matrix in the Yukawa basis, $\widetilde{m}_{\nu} \equiv V_L\,m_{\nu}\,V_L^T$, 
the expressions Eq.~(\ref{MI}) for the RH neutrino masses get generalised as \cite{Akhmedov:2003dg,full}
\be\label{MiVL}
M_1    \simeq   {m^2_{D1} \over |\widetilde{m}_{\nu 11}|} \, , \;\;
M_2  \simeq    {m^2_{D2} \over m_1 \, m_2 \, m_3 } \, {|\widetilde{m}_{\nu 11}| \over |(\widetilde{m}_{\nu}^{-1})_{33}|  } \,  ,  \;\;
M_3  \simeq   m^2_{D3}\,|(\widetilde{m}_{\nu}^{-1})_{33}|  \,  .
\ee
Also the analytical expression Eq.~(\ref{URapp}) for the RH neutrino mixing matrix  
getw nicely generalised as
\be\label{UR}
U_R \simeq  
\left( \begin{array}{ccc}
1 & -{m_{D1}\over m_{D2}} \,  {\widetilde{m}^\star_{\nu 1 2 }\over \widetilde{m}^\star_{\nu 11}}  & 
{m_{D1}\over m_{D3}}\,
{ (\widetilde{m}_{\n}^{-1})^{\star}_{13}\over (\widetilde{m}_{\n}^{-1})^{\star}_{33} }   \\
{m_{D1}\over m_{D2}} \,  {\widetilde{m}_{\nu 12}\over \widetilde{m}_{\nu 11}} & 1 & 
{m_{D2}\over m_{D3}}\, 
{(\widetilde{m}_{\n}^{-1})_{23}^{\star} \over (\widetilde{m}_{\n}^{-1})_{33}^{\star}}  \\
 {m_{D1}\over m_{D3}}\,{\widetilde{m}_{\nu 13}\over \widetilde{m}_{\nu 11}}  & 
- {m_{D2}\over m_{D3}}\, 
 {(\widetilde{m}_\nu^{-1})_{23}\over (\widetilde{m}_\nu^{-1})_{33}} 
  & 1 
\end{array}\right) 
\,  D_{\Phi} \,  ,
\ee
and  the expressions in  Eq.~(\ref{RHphases})  for the three RH neutrino phases as  well:
\be
\Phi_1 = {\rm Arg}[-\widetilde{m}_{\nu 11}^{\star}] \,  , \; \;
\Phi_2 = {\rm Arg}\left[{\widetilde{m}_{\nu 11}\over (\widetilde{m}_{\nu}^{-1})_{33}}\right] -2\,(\rho+\s)-2\,(\rho_L + \s_L) \, , \;  \;
\Phi_3 = {\rm Arg}[-(\widetilde{m}_{\nu}^{-1})_{33}] \,  .
\ee
Starting from these analytical expressions one can also express the flavoured decay parameters and
the flavoured $C\!P$ asymmetries in terms of the parameters in $m_{\nu}$ and $V_L$, explicitly:
\be\label{KialVL}
K_{I\a} = {\sum_{k,l} \, 
m_{Dk}\, m_{Dl} \,V_{L k\a} \, V_{L l \a}^{\star} \, U^{\star}_{R k I} \, U_{R l I} 
\over M_I \, m_{\star}}\,  
\ee
and
\be\label{ve2alAN}
\ve_{2\a} \simeq {3 \over 16\, \pi \, v^2}\,
{|(\widetilde{m}_{\nu})_{11}| \over m_1 \, m_2 \, m_3}\,
{\sum_{k,l} \, m_{D k} \, m_{Dl}  \, {\rm Im}[V_{L k \a }  \,  V^{\star}_{L  l \a } \, 
U^{\star}_{R k 2}\, U_{R l 3} \,U^{\star}_{R 3 2}\,U_{R 3 3}] 
\over |(\widetilde{m}_{\nu}^{-1})_{33}|^{2} + |(\widetilde{m}_{\nu}^{-1})_{23}|^{2}}   \,   .
\ee
Finally, the final $B-L$ asymmetry can be again calculated as the sum of the three contributions
from each flavour,
\be
N_{B-L}^{\rm lep,f} = N_{\D_e}^{\rm lep,f} + N_{\D_\mu}^{\rm lep,f} + N_{\D_\tau}^{\rm lep,f} \,  ,
\ee
where
\bea\label{twofl} \nonumber
N_{\D_e}^{\rm lep,f} & \simeq &
\left[{K_{2e}\over K_{2\tau_2^{\bot}}}\,\ve_{2 \tau_2^{\bot}}\kappa(K_{2 \tau_2^{\bot}}) 
+ \left(\ve_{2e} - {K_{2e}\over K_{2\tau_2^{\bot}}}\, \ve_{2 \tau_2^{\bot}} \right)\,\kappa(K_{2 \tau_2^{\bot}}/2)\right]\,
\, e^{-{3\pi\over 8}\,K_{1 e}}  \,   , \\ \nonumber
N_{\D_\m}^{\rm lep,f} & \simeq & \left[{K_{2\mu}\over K_{2 \tau_2^{\bot}}}\,
\ve_{2 \tau_2^{\bot}}\,\kappa(K_{2 \tau_2^{\bot}}) +
\left(\ve_{2\mu} - {K_{2\mu}\over K_{2\tau_2^{\bot}}}\, \ve_{2 \tau_2^{\bot}} \right)\,
\kappa(K_{2 \tau_2^{\bot}}/2) \right]
\, e^{-{3\pi\over 8}\,K_{1 \mu}} \,  , \\
N_{\D_\t}^{\rm lep,f} & \simeq & \ve_{2 \tau}\,\kappa(K_{2 \tau})\,e^{-{3\pi\over 8}\,K_{1 \tau}} \,  ,
\eea
The second terms in the expression for $N_{\D_e}^{\rm lep,f} $ and $N_{\D_\mu}^{\rm lep,f}$ are
so-called phantom terms \cite{fuller,density}. However, in $SO(10)$-inspired leptogenesis they give
a negligible contribution.  This is because  for all $SO(10)$-inspired solutions one has $K_{2e} \ll K_{2\mu} \simeq K_{2\tau_2^{\bot}}$. 
This means that  the  electronic component of the leptons produced from $N_2$-decays is negligible. Essentially, all leptons produced from $N_2$-decays
have a flavour composition that lies quite precisely on the muon-tauon plane. The reason can be easily understood analytically:
from Eq.~(\ref{KialVL}) and Eq.~(\ref{UR}) one easily derives 
$K_{2e}/K_{2\mu} \lesssim |V_{L21}|^2 \sim \theta^2_{\rm c} \sim 0.05$.
This result, combined with the fact that one also has $\ve_{2e}\ll \ve_{2\mu}$, implies that phantom terms are 
always negligible. 

The final $B-L$ asymmetry, and consequently $\eta_{B0}^{\rm lep}$, now depends also on the 6 parameters in the $V_L$ matrix.
The successful leptogenesis condition will then be now of the form $\eta_{B0}(m_\nu,V_L,\alpha_2) = \eta_{B0}^{\rm obs}$. 
The account of $V_L$ will now generate some thickness to the hypersurface one obtains for $V_L = I$. However,
one still obtains meaningful important constraints on the low energy neutrino parameters. In particular, one still
has a lower bound $m_1 \gtrsim 1\,{\rm meV}$ on the absolute neutrino mass scale. 

Another important aspect of the impact of turning on nonzero angles in $V_L$ is that
the very strong hierarchy of the $N-2$-flavoured $C\!P$ asymmetries in Eq.~(\ref{CPhierarchy}),
now becomes
\be\label{CPhierarchy}
\ve_{2\t}:\ve_{2\m}:\ve_{2e} = \a_3^{\,2}\,m^2_t : \a_2^{\, 2}\,m^2_c = 1 : |V_{L23}| : |V_{L21}|\,|V_{L31}| \,  . 
\ee 
Tauon solutions are still the dominant ones but now muon solutions appear as well.\footnote{In a supersymmetric formulation,
also electronic solutions become viable \cite{susy}.}

A convenient way to display the allowed region is to project it on the three-dimensional space of the 
three neutrino unknown parameters: $m_1, \theta_{23}$ and $\delta$. In Fig~1 we show the result
of a $\sim 2\times 10^9$ trial scan producing a scatter plot of $\sim 2\times 10^6$ solutions for $\alpha_2 =5$
and NO.\footnote{Notice that the success rate is $\sim 0.1\%$. This value should be taken indicatively,
since it depends on the imposed lower limit on the absolute neutrino mass scale that in our case
is $10^{-4}\,{\rm eV}$. Since the scan is done uniformly logarithmically on $m_1$, and since there is a lower bound on $m_1$ on the solutions,
the success rate can be made arbitrarily small decreasing the lower limit on $m_1$.} The solutions
have been obtained imposing:
\begin{itemize}
\item {\small $\chi^2(m_{\rm sol},m_{\rm atm},\theta_{12},\theta_{13}) \equiv 
\left({m_{\rm atm} - \bar{m}_{\rm atm}\over \d m_{\rm atm}}\right)^2 +
\left({m_{\rm sol} - \bar{m}_{\rm sol} \over \d m_{\rm sol}} \right)^2 +
\left({\theta_{12} - \bar{\theta}_{12} \over \d \theta_{12}} \right)^2 +
\left({\theta_{13} - \bar{\theta}_{13} \over \d \theta_{13}} \right)^2
< \chi^2_{\rm max} =25 $;}
\item $\eta_B > \bar{\eta}_B^{\rm exp} - 3\delta \eta_{B}^{\rm exp} = 6.01 \times 10^{-10}$ \,  .
\end{itemize}
\begin{figure}[t]
\centerline{
\psfig{file=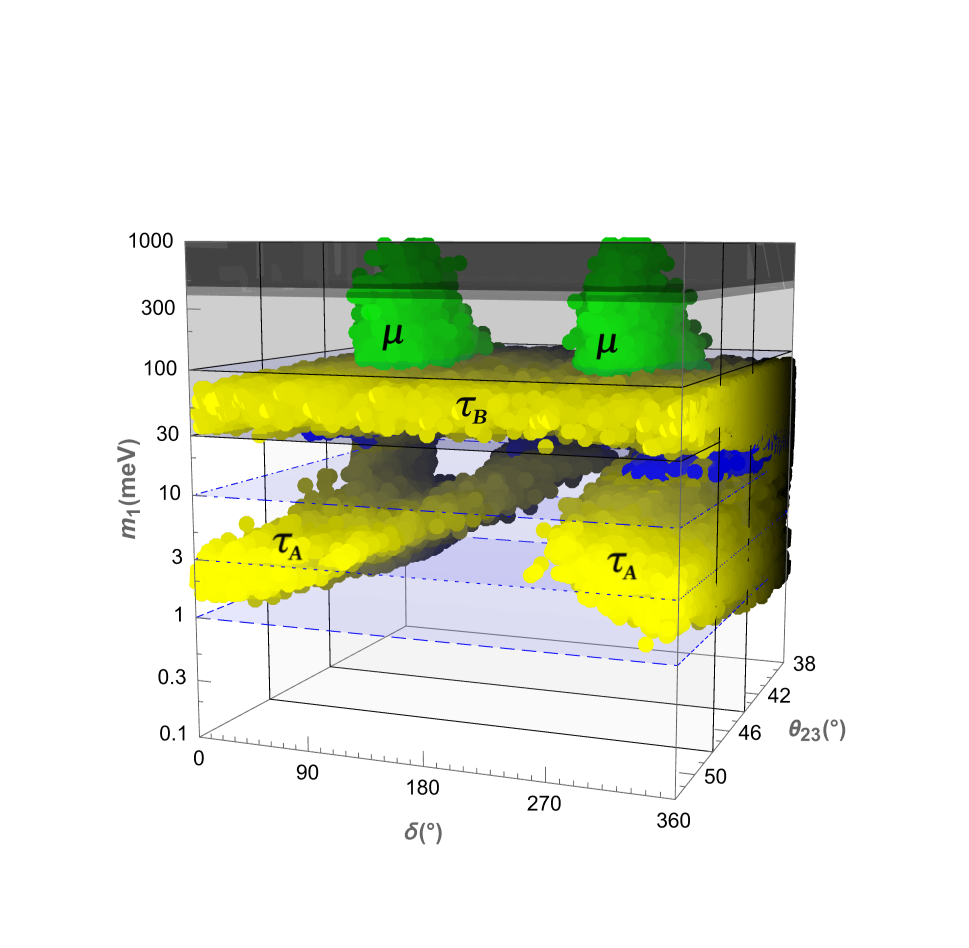,scale=0.5} \hspace{-10mm}
\psfig{file=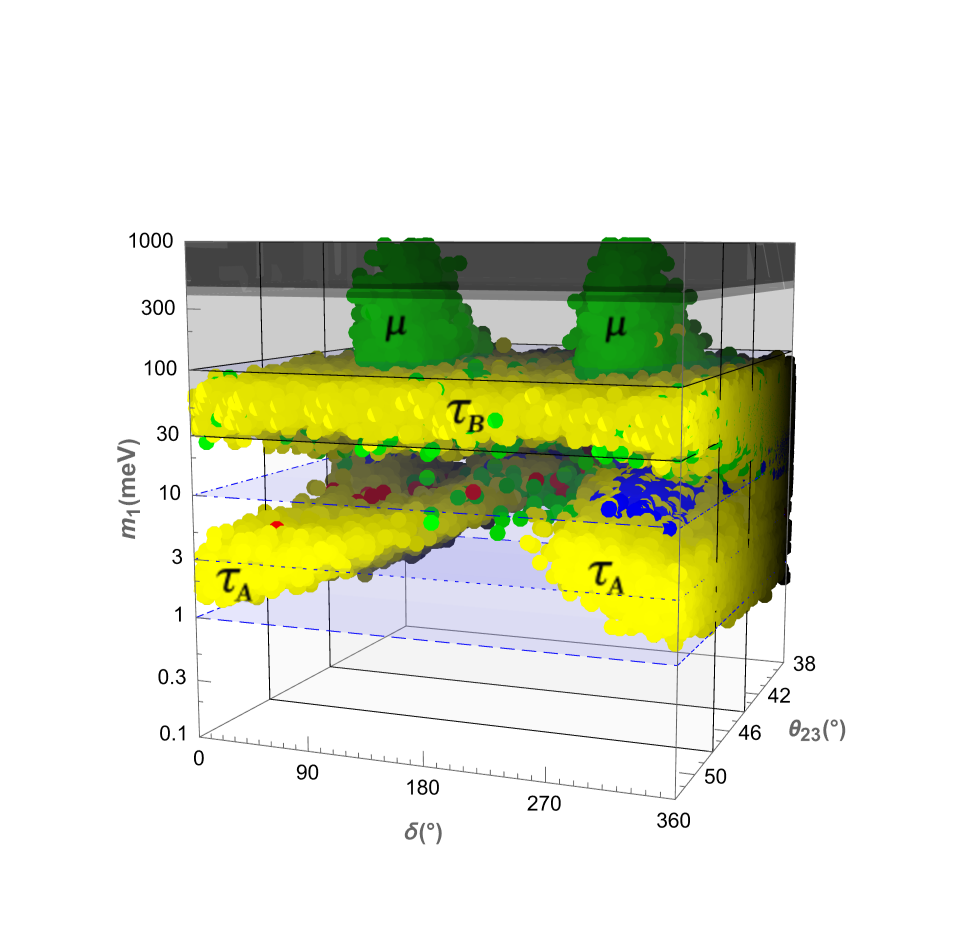,scale=0.5}
}
\caption{Scatter plot of the solutions obtained  imposing successful SO10INLEP neglecting (left) and including (right) 
flavour coupling (from \cite{DiBari:2025zlv}). 
The three grey areas correspond to the excluded regions by the three upper bounds on the absolute neutrino mass scale, in particular, 
Eq.~(\ref{upperbm1}) from cosmological observations and then from $0\nu\b\b$  and from tritium beta decay.  
The three planes in light blue simply help understanding the 3-dim shape.
Colour code: tauonic, muonic and strong thermal solutions are denoted by yellow, green and blue points, respectively.
}
\end{figure}
No solutions are obtained for inverted ordering. This is because these would require too large values of the atmospheric mixing angle
and absolute neutrino mass scale in comparison with current experimental low energy neutrino data that
place upper bounds on both.  This implies that successful $SO(10)$-inspired leptogenesis, with the definition we gave, is incompatible with inverted ordering and, therefore, it predicts normal ordering. We will see soon if this prediction will be successful.

{\em Interplay between absolute neutrino mass scale lower bound and long baseline experiments}. The three-dimensional allowed region shown in Fig.~1 clearly shows the existence of a lower bound on $m_1$ but also that this lower bound is actually depending on the values of $\delta$ and $\theta_{23}$.
In the left panel Fig.~2 one can see the iso-contour lines of the lower bound on $m_1$ in the plane $\delta$ versus $\theta_{23}$.
\begin{figure}[t] \vspace{-1mm}
\psfig{file=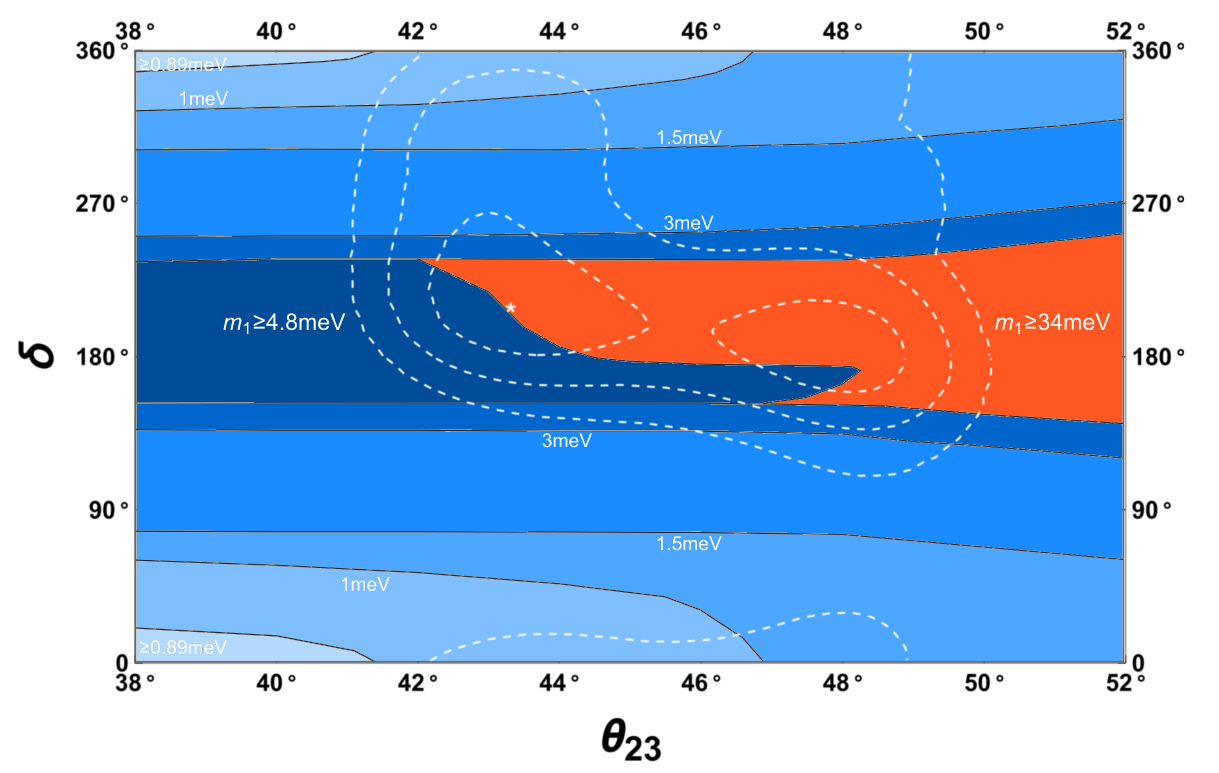,scale=0.4}
\hspace{1mm}
\psfig{file=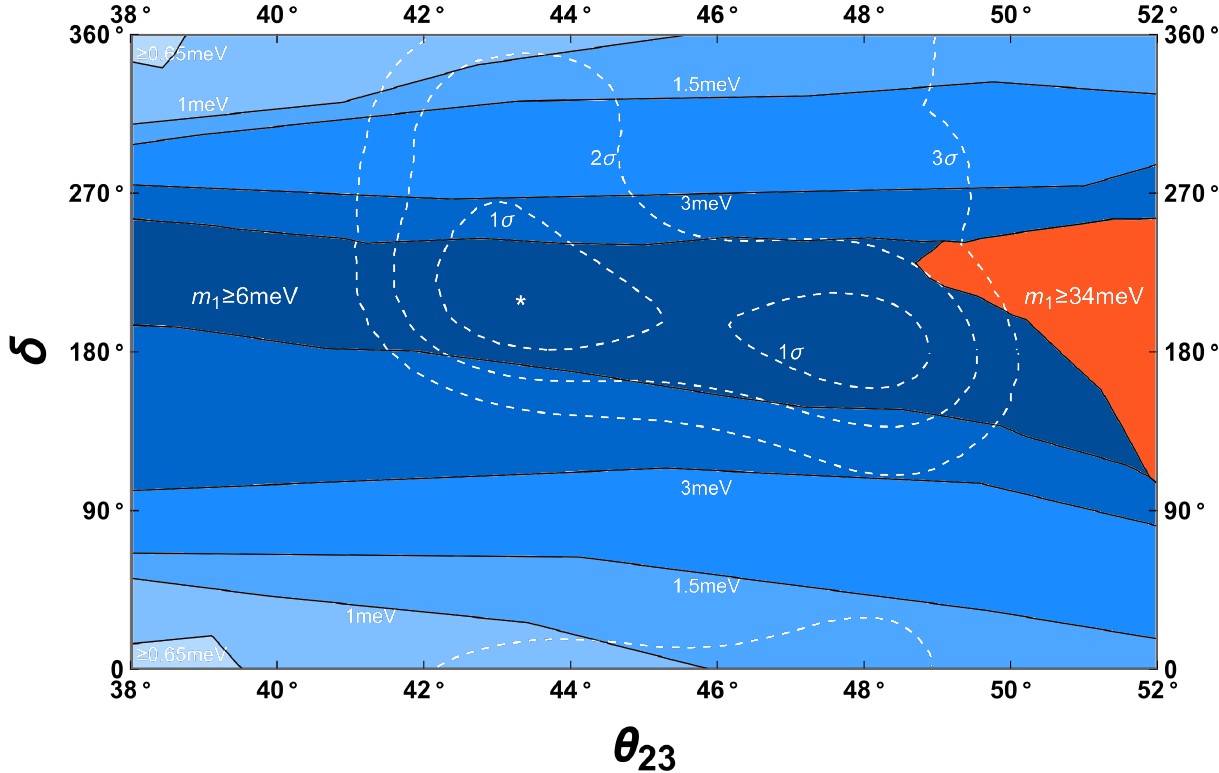,scale=0.4}
\caption{Iso-contour lines of the value of the lower bound on $m_1$ in the $\theta_{23}-\delta$ plane (from \cite{DiBari:2025zlv}) 
without flavour coupling (left) and with flavour coupling (right). The white dashed lines are the $1\sigma$, $2\sigma$and  $3\sigma$ experimental
constraints from \cite{nufit24}. The orange area is the area excluded by the cosmological upper bound (\ref{upperbm1}).}
\end{figure}
One can see how there is a large region (in red) where the lower bound is already incompatible with the cosmological upper bound. This implies that, if future long-baseline experiments will measure $\delta$ and $\theta_{23}$ in this region that $SO(10)$-inspired leptogenesis can be ruled out. However, we will
see how the account of flavour coupling considerably reduces this excluded region.

{\em Upper bound on the atmospheric mixing angle}. Another interesting way to look at the three-dimensional scatter plot in Fig.~1 is to
consider its projection on the $\theta_{23}$ versus $m_1$ plane. The result is shown in the left panel of Fig.~3. 
\begin{figure}[t]
\centerline{
\psfig{file=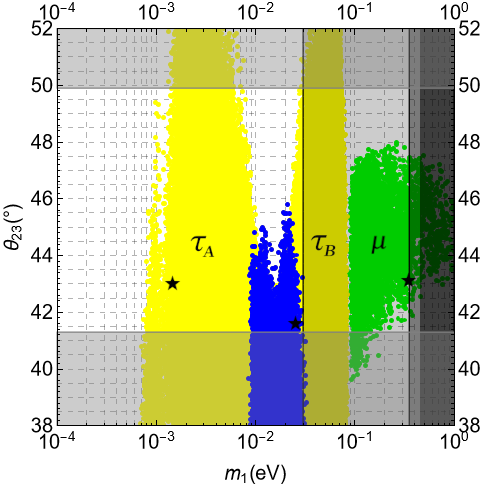,scale=0.5} \hspace{10mm}
\psfig{file=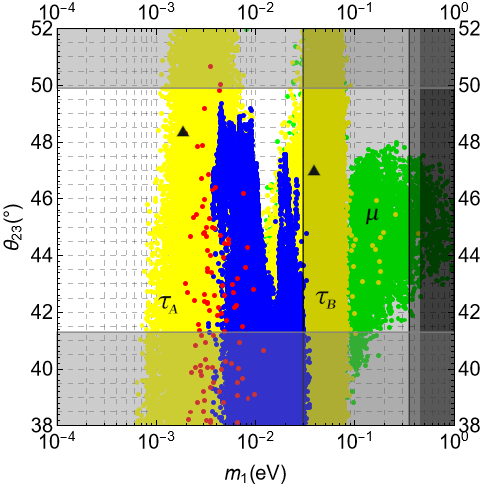,scale=0.5}
}
\caption{Two-dimensional projection on the plane $m_1$-$\theta_{23}$ of the 
scatter plot of the solutions obtained imposing successful SO10INLEP (from \cite{DiBari:2025zlv}), neglecting flavour coupling
effects (left panel)  and accounting for flavour coupling effects (right panel).  Colour code as in Fig.~1.
In the left (right) panel the 3 (2) stars (triangles) denote the 3 (2) some benchmark solutions discussed in detail in \cite{DiBari:2025zlv}.
}
\end{figure}
One can see how in the range $10\,{\rm meV} \lesssim m_1 \lesssim 30\,{\rm meV}$ there are no solutions in the second octant. This implies that
if the atmospheric mixing angle is measured in the second octant, then $SO(10)$-inspired leptogenesis can survive only if $m_1 \lesssim 10\,{\rm meV}$.
Vice-versa, if absolute neutrino mass scale experiments will manage to establish $m_1$ in the range ($10$--$30$) meV, then necessarily the atmospheric micxing angle needs to be in the first octant for successful $SO(10)$-inspired leptogenesis not to be ruled out. This is another interesting
aspect of the interplay between absolute neutrino mass and neutrino mixing unknowns within $SO(10)$-inspired leptogenesis. 

{\em Majorana phases and $0\nu\beta\beta$ effective neutrino mass}. The Majorana phases play a special role in $SO(10)$-inspired leptogenesis
in determining the condition $K_{1\tau}  \lesssim 1$ for tauonic solutions and $K_{1\mu} \lesssim 1$ for muonic solutions. 
For this reason there are strong constraints on Majorana phases imposing successful $SO(10)$-inspired leptogenesis. 
This also interestingly results into the fact that, despite neutrino masses are NO,
there is a lower bound on the effective $0\nu\beta\beta$ neutrino mass $m_{ee}$. In the left panel of Fig.~4 one can see the projection of the 
scatter plot on the plane of Majorana phases, $\rho$ versus $\sigma$, and in the right panel the solutions in the plane $m_{ee}$ versus $m_1$.
\begin{figure}[t]
\centerline{
\psfig{file=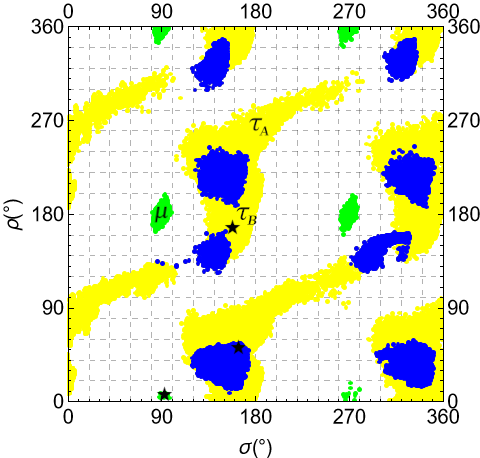,scale=0.5} \hspace{10mm}
\psfig{file=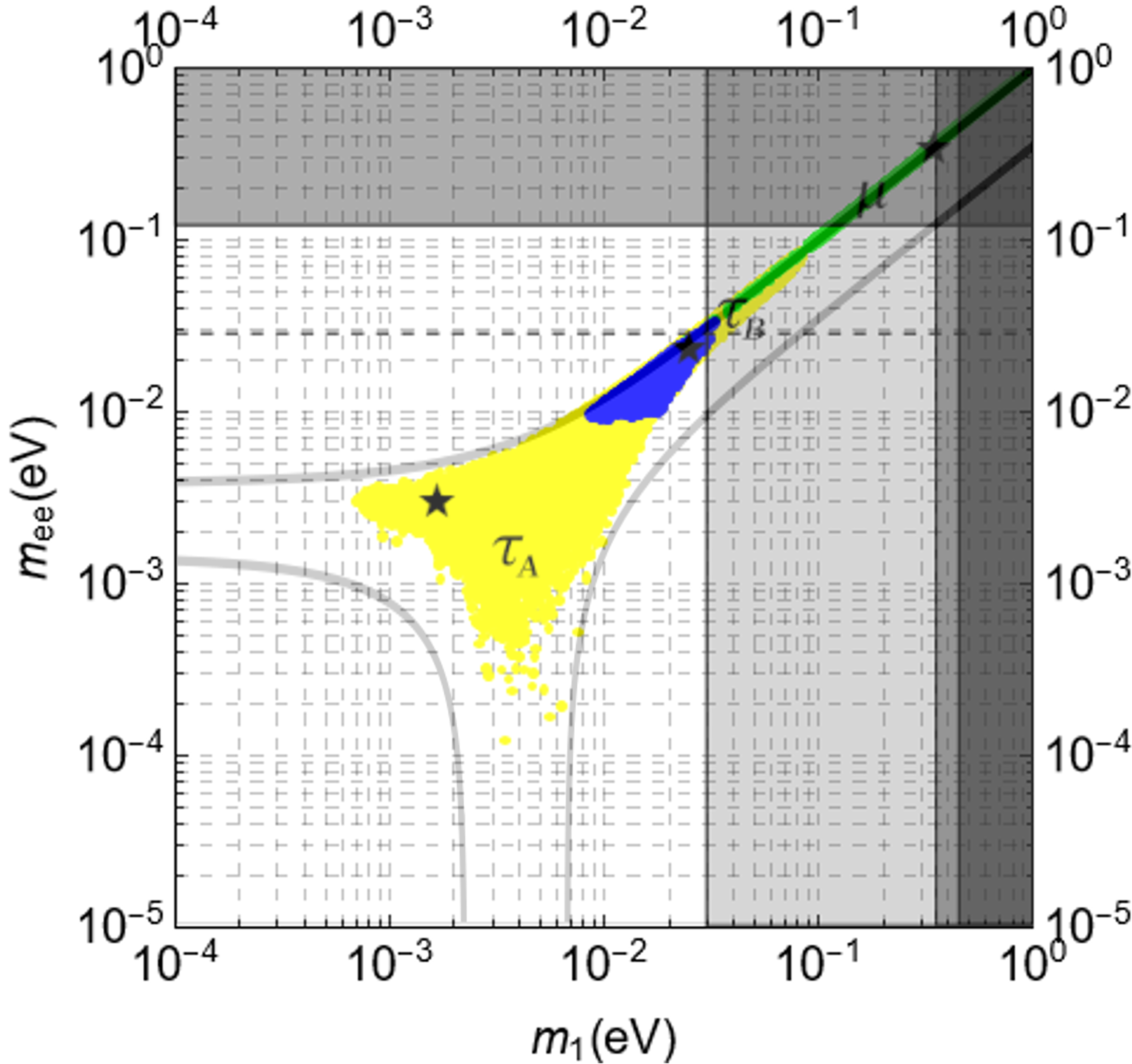,scale=0.39}
}
\caption{Left: Two-dimensional projection on the plane $\sigma$-$\rho$ of the 
scatter plot of the solutions obtained imposing successful SO10INLEP.
Right: Solutions in the plane $m_{ee}$ versus $m_1$. 
}
\end{figure}

\section{Strong thermal $SO(10)$-inspired leptogenesis}

As we mentioned, $N_2$-leptogenesis is the only scenario that can realise strong thermal leptogenesis for a hierarchical RH neutrino mass
spectrum when flavour effects are taken into account. However, there are a few additional conditions that need to be
satisfied \cite{Bertuzzo:2010et}. An initial large pre-existing asymmetry $N^{\rm p,i}_{B-L}$ has to be washed-out first in the tauon flavour by $N_2$-inverse processes in the two-flavour regime and this requires $K_{2\tau} \gg 1$. The electron and muon flavour components of the pre-existing asymmetry have then to be
washed out by $N_1$-inverse processes in the three-flavoured regime and this requires $K_{1\mu}, K_{1e} \gg 1$. On the other hand, the asymmetry
produced by $N_2$-decays in the tauon flavour has to survive the $N_1$-washout and this requires additionally $K_{1\tau} \lesssim 1$. 
For all these conditions to be satisfied, necessarily $m_1 \gtrsim 10\,{\rm meV}$ for $N^{\rm p,i}_{B-L}\sim 10^{-3}$ (there is a logarithmic dependence
of the lower bound on $N^{\rm p,i}_{B-L}$) \cite{DiBari:2014eqa}.    

Apparently, this is such a special set of conditions that it would seem quite difficult to find realistic models satisfying them. For this reason, it
is highly non trivial that all these conditions are satisfied by a subset of the successful $SO(10)$-inspired leptogenesis solutions we 
showed \cite{ST}. This subset is denoted by the blue colour in Fig.~1, Fig.~3 and Fig.~4. They are tauonic solutions with $K_{2\tau} \gg 1$ (indicated as $\tau_A$
solutions in the plots) and with $m_1 \gtrsim 10\,{\rm meV}$ (we used $N^{\rm p,i}_{B-L}\sim 10^{-3}$). 

Clearly the constraints on the low energy neutrino parameters become quite stringent,  yielding well definite predictions. 
In Fig.~5 we show the results of the original search of solutions of strong thermal $SO(10)$-insired leptogenesis \cite{ST}
projects on three different planes of low energy neutrino parameters. 
\begin{figure}[t]
\centerline{
\psfig{file=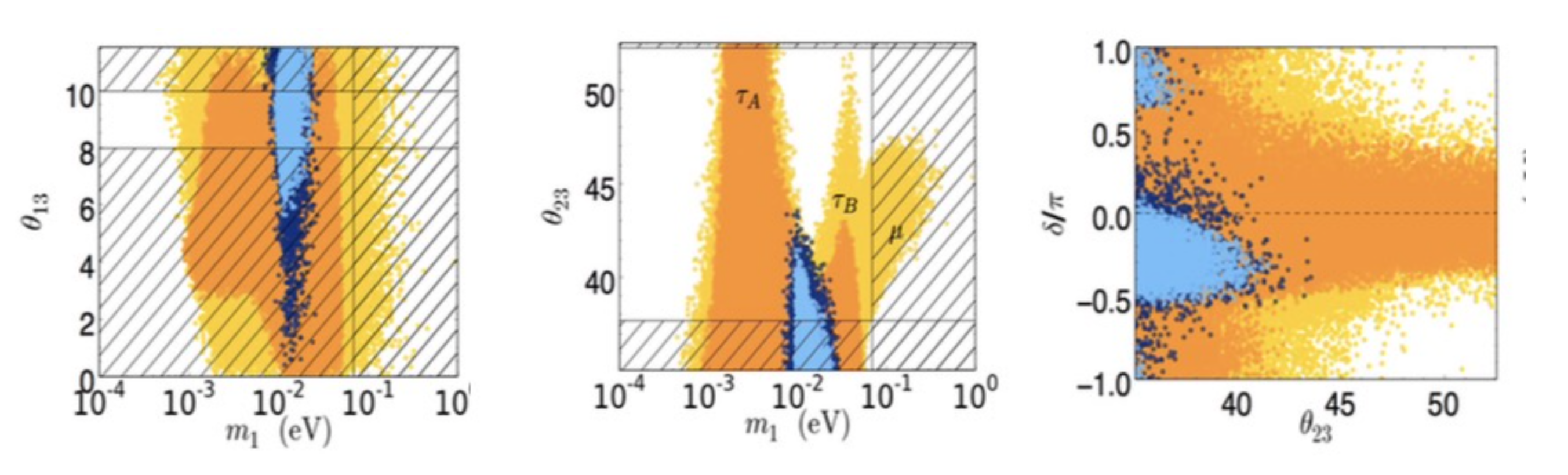,scale=0.35} 
}
\caption{In these three panels, one can see how strong thermal $SO(10)$-inspired leptogenesis: 
(i) it requires a non-vanishing $\theta_{13}$ (left);
(ii) it makes a sharp prediction on $m_1$ and implies a stringent upper bound on $\theta_{23}$ (centre and right); 
(iii) it strongly favours $\delta$ in the fourth quadrant (right).
}
\end{figure}
These three panels show well the 
three important features of the strong thermal $SO(10)$-inspired leptogenesis solution:
\begin{itemize}
\item In the left panel one can see how strong thermal $SO(10)$-inspired leptogenesis has successfully {\bf pre}dicted a non-vanishing $\theta_{13}$.
Indeed, first results were presented in September 2011 at the DESY theory workshop \cite{preliminary} before the DayaBay discovery 
in January 2012 \cite{DayaBay:2012fng}. This is a remarkable success for this scenario of leptogenesis.
\item From the central panel, one can see how the solutions are confined within the range $10\,{\rm meV} \lesssim  m_1 \lesssim 30\,{\rm meV}$ 
and, for what we said, this implies  a stringent upper bound on $\theta_{23}$.  
\item It strongly favours $\delta$ in the fourth quadrant. 
\end{itemize}
In addition to these three features, it is very interesting that the lower bound $m_1 \gtrsim 10\,{\rm meV}$  implies an analogous
lower bound on the effective $0\nu\beta\beta$ neutrino effective mass $m_{ee} \gtrsim 10\,{\rm meV}$ (see right panel in Fig.~4). It is intriguing that the current range of values, accounting for nuclear matrix
elements uncertainties,  for the experimental upper bound on $m_{ee}$ from the KamLAND-Zen experiment \cite{KamLAND-Zen:2024eml}
(shown in the right panel of Fig.~4 as the range comprised between the horizontal dashed  line for $m_{ee}\simeq 30\,{\rm meV}$
and the horizontal solid line for $m_{ee} \simeq 120\,{\rm meV}$) starts to overlap with the range of values
compatible with strong thermal $SO(10)$-inspired leptogenesis. This shows that despite the solutions exist only for NO, 
still strong thermal $SO(10)$-inspired leptogenesis predicts that a $0\nu\beta\beta$ signal should be detected during approximately the 
next decade. In particular,  KamLAND2-Zen aims at starting in 2027 and improving the upper bound to $m_{ee} \lesssim 20\,{\rm meV}$ 
within  5 years \cite{shimizu}. 

{\em Strong thermal $SO(10)$-inspired leptogenesis confronting long baseline experiments.} Let us analyse in greater detail 
the theoretical predictions with the experimental results on the two neutrino oscillation unknowns: atmospheric mixing angle $\theta_{23}$
and $C\!P$ violating phase $\delta$. A dedicated analysis has been conducted more accurately to determine the upper bound on $\theta_{23}$
as a function of $\delta$ \cite{chianese}. The results are shown in the left panel of Fig.~6.
\begin{figure}[t]
\centerline{
\psfig{file=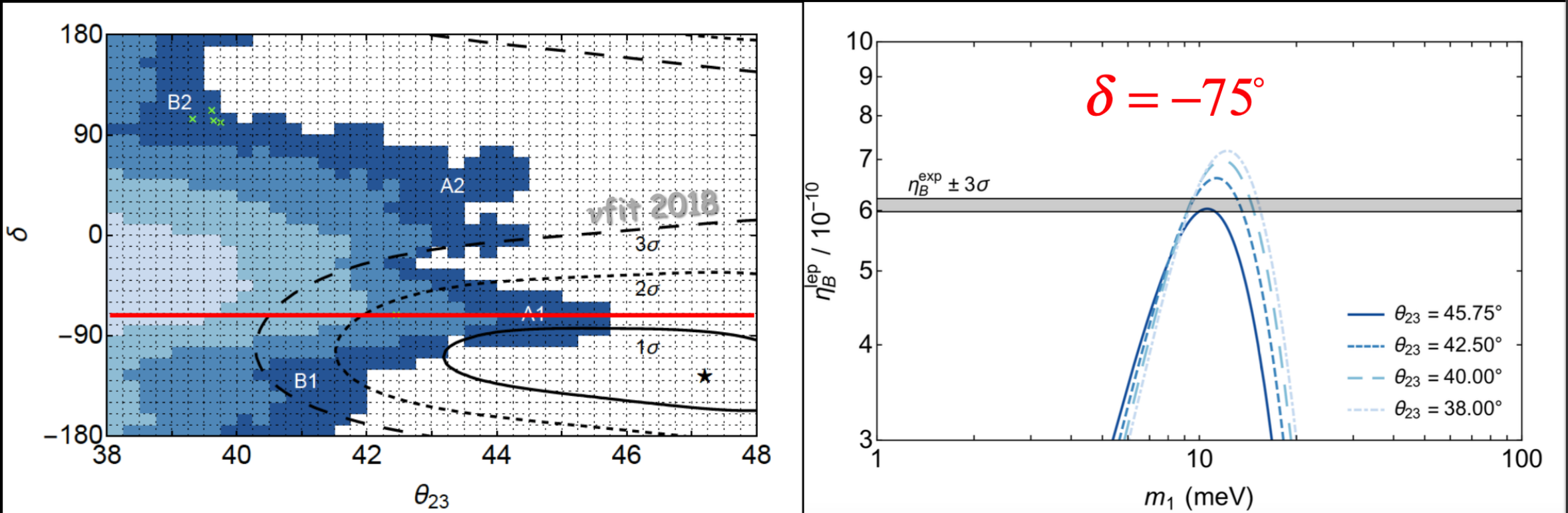,scale=0.33} 
}
\caption{Left: determination of the upper bound on $\theta_{23}$ as a function of $\delta$.  
Right: Value of the asymmetry for fixed $\delta = - 75^\circ$ and for the four indicated values of $\theta_{23}$ as a function of $m_1$ (from \cite{chianese}).}
\end{figure}
In this case the scatter plot also retains information on the density of points, indicated by the colour code (from light to dark blue, density is decreasing).
One can see how the bulk of solutions is contained in the fourth quadrant of $\delta$. The maximum value of $\theta_{23}$ is $\simeq 46^\circ$
and is saturated for $\delta \simeq -75^\circ$.  In the right panel one can see  plots of the $\eta_{B0}^{\rm lep}$ versus $m_1$
for four different values of $\theta_{23}$. It clearly shows how the asymmetry is suppressed for increasing value of $\theta_{23}$, 
the origin of the upper bound.  In the left panel one can also see the experimental constraints on the plane $\theta_{23}$--$\delta$
from $\nu$fit global analysis in 2018 \cite{Esteban:2018azc}. At that time, data were clearly favouring $\theta_{23}$ in the second octant.
For this reason  the  strong thermal $SO(10)$-inspired leptogenesis solution seemed to start to be cornered. 

{\em New atmospheric neutrino data seem to remove the tension}. However, when new data from atmospheric neutrinos \cite{Super-Kamiokande:2023ahc}  are implemented,  global analyses find now the best fit for $\theta_{23}$ in the first octant \cite{nufit24}.   
A comparison between these new experimental constraints and the strong thermal $SO(10)$-inspired leptogenesis allowed region in the $\delta$ versus $\theta_{23}$ plane can be seen in Fig.~7 (in this case one has to look at the region at the left of the red line, the blue regions include flavour coupling that we will discuss in the next section). 
\begin{figure}[t]
\centerline{
\psfig{file=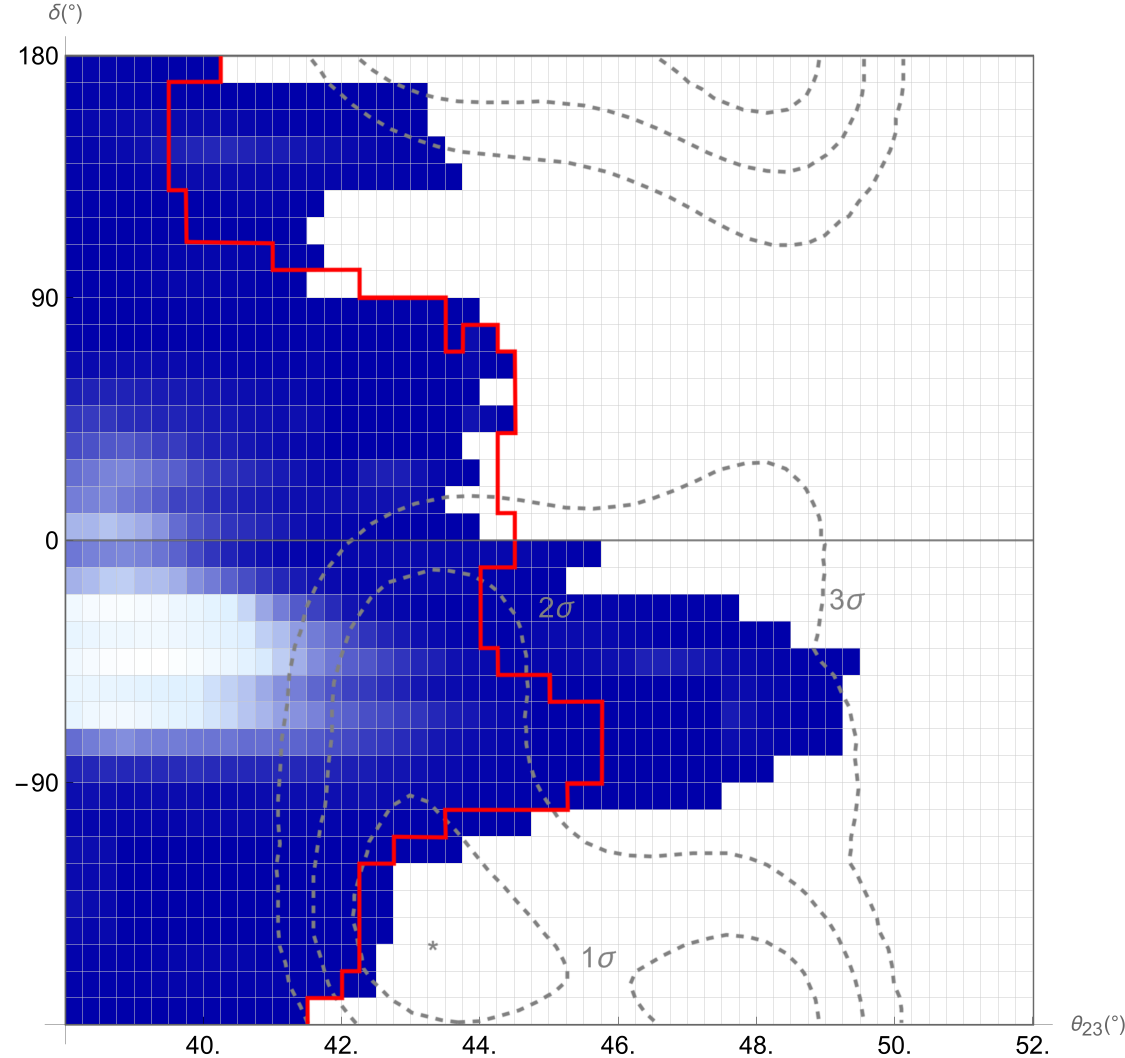,scale=0.33} 
}
\caption{Strong thermal $SO(10)$-inspired leptogenesis allowed region in the plane $\delta$ versus $\theta_{23}$. The red line  delimits the region when 
flavour coupling is neglected, the coloured region is the result in the case when flavour coupling is taken into account. The colour code
denotes point density:  lighter blue means higher density of solutions. The thin dashed lines are
$1\sigma, 2\sigma$ and $3\sigma$ experimental constraints from \cite{nufit24} (figure taken from \cite{DiBari:2025zlv}).}
\end{figure}
 One can see how now the agreement is quite good and there is overlap even at $1\sigma$. In any case
 it is clear that the experimental determination of $\theta_{23}$ and $\delta$ is currently 
 subject to many uncertainties and instabilities. We need to wait for DUNE and T2HK for a more precise and accurate 
 determination of $\theta_{23}$ and $\delta$ with long baseline experiments \cite{Ballett:2016daj} that will
 provide an ultimate test of the strong thermal $SO(10)$-inspired leptogenesis solution.
 
\section{Impact of flavour coupling}

{\em Theoretical uncertainties.}  There are four main effects that are neglected in the calculation of the asymmetry 
that we discussed so far:
\begin{itemize}
\item Flavour coupling from spectator processes;
\item Radiative corrections and running of parameters;
\item Full density matrix calculation;
\item Momentum dependence. 
\end{itemize}
An account of each of these effects is expected to give some correction to the results we discussed so far. 
At the same time their account slows down the determination of the allowed region in the space of parameters 
requiring scatter plots with millions of points (needing billions of trials). Fortunately, the discussed analytical derivation
of the RH neutrino masses and mixing matrix facilitates this task. 

Out of these four effects, flavour coupling is expected to give the largest correction to the calculation of the
asymmetry. Moreover, in principle, one could wonder whether an account of flavour coupling could jeopardise the
strong thermal solution, since the pre-existing asymmetry could leak from the strongly washed-out electron and muon flavours 
to the tauon flavour at the lightest RH neutrino scale. For these reasons, it is  the effect with highest priority to be included.

{\em Including flavour coupling.}

We have so far assumed that the evolution of the asymmetry in each flavour is independent of each other and, therefore,
described by an uncoupled set of three differential equations. However, the account of spectator processes \cite{Buchmuller:2001sr,Barbieri:1999ma,Nardi:2005hs}, 
dominantly the Higgs asymmetry \cite{bounds}, will generate a coupling in the evolution of the flavoured asymmetries that will be, therefore, described by
a coupled set of differential equations both at the $N_2$-production and at the $N_1$-washout. 
In particular, at the $N_1$-stage, one has a set of coupled differential equations of the form
\be\label{flkewA}
{dN_{\D_{\a}}\over dz_1}  =
-{K_{1\a}\over K_1}\,\sum_{\b}\,C^{(3)}_{\a\b}\,W_1^{\rm ID}\,N_{\D_{\b}} \, ,
\hspace{15mm} (\a,\b=e,\m,\t)
\ee
where $C^{(3)}_{\a\b}$ is the flavour coupling matrix in the three-flavour regime (a detailed discussion can be found in \cite{fuller}).
Solving this system of coupled differential equations, one finds that each final flavoured asymmetry is now given by an expression of the form:
\bea\label{NfDa2}
N^{\rm f}_{\D_{\a}}
 & = &  V^{-1}_{\a e''}\,
\left[\sum_{\b}\,V_{e''\b}\,N_{\D_{\b}}^{T\sim T_L}\right]
\,e^{-{3\pi\over 8}\,K_{1 e''}} \\  \nonumber
& + &  V^{-1}_{\a \m''}\,
\left[\sum_{\b}\,V_{\m''\b}\,N_{\D_{\b}}^{T\sim T_L}\right]
\,e^{-{3\pi\over 8}\,K_{1 \m''}} \\ \nonumber
& + &  V^{-1}_{\a \t''}\,
\left[\sum_{\b}\,V_{\t''\b}\,N_{\D_{\b}}^{T\sim T_L}\right]
\,e^{-{3\pi\over 8}\,K_{1 \t''}} \,  ,
\eea
where $V$ is a matrix diagonalising $(K_{1\alpha}/K_1)\,C^{(3)}_{\a\b}$ and $V^{-1}$ its inverse. 
One can see that each final flavoured asymmetry is now the sum of 9 terms, so that the final $B-L$
asymmetry is now given by the sum of 27 terms instead of just one as in the uncoupled case. This clearly shows
how the generation of a scatter plot of million points  becomes much more computing demanding and 
analytical expressions greatly help such a task that was carried out in  \cite{DiBari:2025zlv}.  
The result is shown in the right panel of Fig.~1.  One can see how the main change is the appearance of new muonic solutions.
These muonic solutions are intrinsically different from those obtained neglecting flavour coupling. 
The asymmetry is in the first place produced in the tauon flavour but then, thanks to flavour coupling, some fractions leaks
into the muon flavour. The interesting aspect of these solutions is that they appear in a region of the $\delta$ versus $\theta_{23}$
plane that would otherwise be unaccessible.  This can be clearly seen in the right panel of Fig.~2. One can see how now muonic solutions with low value
of $m_1$ compatible with the cosmological upper bound on neutrino masses, appear in a region of the $\delta$ versus $\theta_{23}$ plane where there were no solutions before.  However, it should be said how these muonic solutions are sub-dominant compared to the standard tauonic solutions, they represent only
$1\%$ of the total number of solutions. There are even more special electronic solutions appearing (in red colour). However, these solutions overlap with the already existing tauonic solutions in the parameter space.

Finally, one can consider the impact of flavour coupling on the strong thermal $SO(10)$-inspired leptogenesis solution. 
The first question to answer is whether flavour coupling might completely jeopardise the solution. However, the answer is fortunately negative
and flavour coupling just simple introduces some correction. The results are given by the blue regions in Fig.~7 to be compared with the 
region at the left of the red line obtained neglecting flavour coupling. One can see how the inclusion of flavour coupling even enlarges the region relaxing the upper bound on the atmospheric mixing angle. However, the bulk region of the solutions is still pretty much unchanged. It is interesting to understand why
flavour coupling does not jeopardise the solution. One could suspect that the pre-existing asymmetry could 
survive the $N_1$-washout leaking from the electron and muon flavours, where the washout is strong, to the tauon flavour, where the 
washout is weak. This does not happen simply because the same weak tauonic washout, $K_{1\tau}\lesssim 1$, reduces to the coupling of the tauon flavour
at the $N_1$-washout stage: the solution kind of self-protects itslef by this flavour coupling potential threat. This is a very encouraging 
feature that shows an intrinsic stability of the solution.

\section{A recent realistic fit} 

The last aspect I want to discuss is whether $SO(10)$-inspired leptogenesis can be realised within some realistic model. Recently, it  was shown
that a good fit of fermion parameters  can be obtained within a minimal (renormalisable) $SO(10)$ model \cite{Babu:2016bmy} and this can also reproduce the observed baryon asymmetry with $N_2$-leptogenesis \cite{Babu:2024ahk}.  The Higgs fields belong to (real) 10-dim, (real) 120-dim and (complex) 126-dim. representations and have Yukawa couplings with three chiral families of quarks and leptons. Each family is unified into a single irreducible representation 
of $SO(10)$ given by a 16-dimensional spinor which also contains one RH neutrino.  After $SO(10)$ SSB, fermion mass matrices are generated with a number of parameters lower that in the SM. Therefore, it is not trivial that a fit to observed fermion masses and mixing parameters can be found. It is even less trivial that a fit also to the baryon asymmetry generated with $N_2$-leptogenesis can be found. We refer to \cite{Babu:2016bmy} for a table of all values of the parameters of the fit. Here we just focus on the parameters relevant for a comparison with the results of $SO(10)$-inspired leptogenesis we discussed.

First of all the RH neutrino mass spectrum is just a specific realisation of the general spectrum we have seen
to hold in $SO(10)$-inspired models: $M_1 \simeq 7\times 10^4\,{\rm GeV}$, $M_2 \simeq 2 \times 10^{12}\,{\rm GeV}$
and $M_3 \simeq 8 \times 10^{14}\,{\rm GeV}$.  Clearly, the asymmetry requires $N_2$-leptogenesis. 
The light neutrino masses are NO.\footnote{We also find a fit with IO neutrino masses but the goodness of the fit is much worse (much higher $\chi^2$). However, it is interesting to notice
that violating the $SO(10)$-inspired condition $V_L \lesssim V_{CKM}$, IO solutions become less disfavoured.} 
However, one has $m_1 \simeq 0.038\,{\rm meV}$
and, consequently, $m_{ee}\simeq 3.7\,{\rm meV}$. This clearly violates the lower bound $m_1 \gtrsim 1\,{\rm meV}$
we discussed in detail.  

An inspection of the Dirac neutrino mass matrix from the fit shows that the angle $\theta_{23}^L$ in the $V_L$ matrix is maximal.
This clearly indicates a deviation from the $SO(10)$-inspired conditions.  It was indeed already noticed in \cite{DiBari:2020plh} that increasing
$\theta_{23}^L$ enhances the muonic $C\!P$ asymmetry extending the allowed region for muon solutions to smaller $m_1$ values.
In this way the lower bound on $m_1$ can be relaxed. This would suggest that $SO(10)$-inspired leptogenesis can be extended strongly violating the conditions 
$V_L \lesssim V_{CKM}$ but in a way that $N_2$-leptogenesis still holds but with an extended region of viability. It should also be 
clarified that the fit we discussed does not realises strong thermal leptogenesis (in that case the lower bound $m_1\gtrsim 10\,{\rm meV}$ is necessary).  

A more extensive numerical search has recently found more points within this minimal $SO(10)$ model \cite{Chen:2025afg} with solutions also in the IO case (though a much more reduced set). However, it is not possible to map these results with our approach, since the biunitary parameterisation is not shown. Likely, in these fits the angles in $V_L$ are allowed to be arbitrarily large, 
since in this case IO solutions are indeed possible.   

Recently, it was also emphasized the role that a triplet Higgs might have
in $SO(10)$-leptogenesis and in the search of realistic minimal $SO(10)$-models \cite{Fong:2025aya}.

\section{Conclusions}

\begin{itemize}
\item $SO(10)$-inspired leptogenesis is a scenario of leptogenesis relying on $N_2$-leptogenesis
and is well motivated within grandunified models (not necessarily $SO(10)$ models, though these are certainly the most
attractive candidates to realise it).   It leads to interesting predictions, in particular a non-vanishing value of the absolute 
neutrino mass scale, $m_1\gtrsim 10\,{\rm meV}$, and, normally ordered neutrino masses. 
The latter is quite timely to be highlighted in view of the imminent results expected 
from the JUNO experiment \cite{JUNO:2025gmd}. 
\item The atmospheric neutrino mixing angle is predicted in the first octant if 
$10\,{\rm meV} \lesssim m_1 \lesssim 30\,{\rm meV}$, a range of values that now start to be tested cosmologically. 
\item A subset of the solutions realises strong thermal leptogenesis. This implies of course even more stringent predictive constraints on the low energy neutrino parameters.
The atmospheric neutrino mixing angle has to be strictly in the first octant and $\delta$ is preferred in the fourth quadrant. 
The lower bound on the absolute neutrino mass scale becomes $m_1 \gtrsim 10\,{\rm meV}$ and, very interestingly,
this also results into an analogous lower bound on the effective $0\nu\beta\beta$ neutrino mass.  Ultimately, the detection
of $0\nu\beta\beta$ decay signal with such a value of $m_{ee}$, if accompanied by the success of the other predictions, would likely
provide a way to claim a strong evidence, if not a discovery, of the strong thermal $SO(10)$inspired leptogenesis solution. 
\item An account of flavour coupling introduces new muonic solutions but it does not change the overall picture and, importantly, it 
does not jeopardise the strong thermal solution.
\item $SO(10)$-inspired leptogenesis can be realised within a minimal $SO(10)$ model, though this requires an extension of the 
$SO(10)$-inspired conditions, where the atmospheric-like mixing angle in the $V_L$ matrix has to be close-to-maximal ($\theta_{23}^L \simeq 45^\circ$, something that might suggest the  presence of a discrete flavour symmetry in addition to $SO(10)$).
\end{itemize}
It is an exciting time for the prospects to test $SO(10)$-inspired leptogenesis in light of
new expected results from low energy neutrino experiments during next years, both from long baseline and $0\nu\beta\beta$ experiments. 

\subsubsection*{Acknowledgments}
I acknowledge financial support from the STFC Consolidated Grant ST/T000775/1.
I also acknowledge support from the European Union’s Horizon 2020 Europe research and innovation programme under  
the Marie Sk\l odowska-Curie grant agreement HIDDeN European  ITN project (H2020-MSCA-ITN2019//860881-HIDDeN). \\
It is a pleasure to thank Kaladi Babu, Enrico Bertuzzo, Marco Chianese, Chee Sheng Fong, Sophie King, Steve King, Xubin Hu, Michele Re Fiorentin, Luca Marzola, 
Toni Riotto, Shaikh Saad, Rome Samanta, for a fruitful collaboration on $SO(10)$-inspired leptogenesis and related subjects.


\begin{thebibliography}{99}

\bibitem{DiBari:2025zlv}
P.~Di Bari and X.~Hu,
{\em Impact of flavour coupling on SO(10)-inspired leptogenesis},
JCAP \textbf{01} (2026), 024
[arXiv:2507.06144 [hep-ph]].

\bibitem{Esteban:2026phq}
I.~Esteban, M.~C.~Gonzalez-Garcia, M.~Maltoni, I.~Martinez-Soler, J.~P.~Pinheiro and T.~Schwetz,
{\em Lessons from the first JUNO results},
[arXiv:2601.09791 [hep-ph]].

\bibitem{JUNO:2025gmd}
A.~Abusleme \textit{et al.} [JUNO],
{\em First measurement of reactor neutrino oscillations at JUNO},
[arXiv:2511.14593 [hep-ex]].

\bibitem{Tristram:2023haj}
M.~Tristram, A.~J.~Banday, M.~Douspis, X.~Garrido, K.~M.~G{\'o}rski, S.~Henrot-Versill{\'e}, L.~T.~Hergt, S.~Ili{\'c}, R.~Keskitalo and G.~Lagache, \textit{et al.}
{\em Cosmological parameters derived from the final Planck data release (PR4)},
Astron. Astrophys. \textbf{682} (2024), A37
[arXiv:2309.10034 [astro-ph.CO]].

\bibitem{KamLAND-Zen:2024eml}
S.~Abe \textit{et al.} [KamLAND-Zen],
{\em Search for Majorana Neutrinos with the Complete KamLAND-Zen Dataset},
[arXiv:2406.11438 [hep-ex]].

\bibitem{Katrin:2024tvg}
M.~Aker \textit{et al.} [Katrin],
{\em Direct neutrino-mass measurement based on 259 days of KATRIN data},
[arXiv:2406.13516 [nucl-ex]].


\bibitem{nufit24}
I.~Esteban, M.~C.~Gonzalez-Garcia, M.~Maltoni, I.~Martinez-Soler, J.~P.~Pinheiro and T.~Schwetz,
{\em NuFit-6.0: Updated global analysis of three-flavor neutrino oscillations},
[arXiv:2410.05380 [hep-ph]].
  
\bibitem{seesaw}
P.~Minkowski,
  {\em $\mu\to e \g$ At A Rate Of One Out Of 1-Billion Muon Decays?},
  Phys.\ Lett.\  B {\bf 67} (1977) 421;
T. Yanagida, 
{\em Horizontal gauge symmetry and masses of neutrinos},
  Conf.\ Proc.\ C {\bf 7902131} (1979) 95.
 Proceedings of the Workshop on Unified Theory and Baryon Number
of the Universe, eds. O. Sawada and A. Sugamoto (KEK, 1979) p.95;
M.~Gell{\nobreakdash-}Mann, P.~Ramond and R.~Slansky,
{\em The Family Group in Grand Unified Theories},
in Sanibel Conference (Feb 1979), CALT-68-709, hep-ph/9809459
and {\em Complex Spinors and Unified Theories}, in Supergravity, 
PRINT-80-0576, Conf. Proc. C \textbf{790927} (1979), 315-321 [arXiv:1306.4669 [hep-th]];
 R.~Barbieri, D.~V.~Nanopoulos, G.~Morchio and F.~Strocchi,
  {\em Neutrino Masses in Grand Unified Theories},
  Phys.\ Lett.\  {\bf 90B} (1980) 91.
R.~N.~Mohapatra and G.~Senjanovic,
  {\em Neutrino Mass and Spontaneous Parity Nonconservation},
  Phys.\ Rev.\ Lett.\  {\bf 44} (1980) 912.
  
\bibitem{Altarelli:2010gt}
G.~Altarelli and F.~Feruglio,
{\em Discrete Flavor Symmetries and Models of Neutrino Mixing},
Rev. Mod. Phys. \textbf{82} (2010), 2701-2729
[arXiv:1002.0211 [hep-ph]].
  
\bibitem{King:2013eh}
S.~F.~King and C.~Luhn,
{\em Neutrino Mass and Mixing with Discrete Symmetry},
Rept. Prog. Phys. \textbf{76} (2013), 056201
[arXiv:1301.1340 [hep-ph]].
  
\bibitem{Bertuzzo:2009im}
E.~Bertuzzo, P.~Di Bari, F.~Feruglio and E.~Nardi,
{\em Flavor symmetries, leptogenesis and the absolute neutrino mass scale},
JHEP \textbf{11} (2009), 036
[arXiv:0908.0161 [hep-ph]].
  
\bibitem{DiBari:2018fvo}
P.~Di Bari, M.~Re Fiorentin and R.~Samanta,
{\em Representing seesaw neutrino models and their motion in lepton flavour space},
JHEP \textbf{05} (2019), 011
[arXiv:1812.07720 [hep-ph]].

\bibitem{Akhmedov:2003dg}  
E.~K.~Akhmedov, M.~Frigerio and A.~Y.~Smirnov,
{\em Probing the seesaw mechanism with neutrino data and leptogenesis},
JHEP \textbf{09} (2003), 021
[arXiv:hep-ph/0305322 [hep-ph]].

\bibitem{decrypting}
P.~Di Bari, L.~Marzola and M.~Re Fiorentin,
  {\em Decrypting $SO(10)$-inspired leptogenesis},
  Nucl.\ Phys.\ B {\bf 893} (2015) 122
  [arXiv:1411.5478 [hep-ph]].

\bibitem{SO10inspired}
A.~Y.~Smirnov,
 {\em Seesaw enhancement of lepton mixing},
  Phys.\ Rev.\ D {\bf 48} (1993) 3264
  [hep-ph/9304205];
W.~Buchmuller and M.~Plumacher,
  {\em Baryon asymmetry and neutrino mixing},
  Phys.\ Lett.\ B {\bf 389} (1996) 73 [hep-ph/9608308];
F.~Buccella, D.~Falcone and F.~Tramontano,
  {\em Baryogenesis via leptogenesis in SO(10) models},
  Phys.\ Lett.\ B {\bf 524} (2002) 241 [hep-ph/0108172];
   
 \bibitem{orloff}
E.~Nezri and J.~Orloff,
{\em Neutrino oscillations versus leptogenesis in SO(10) models},
JHEP \textbf{04} (2003), 020
[arXiv:hep-ph/0004227 [hep-ph]].
   
\bibitem{Branco:2002kt}
G.~C.~Branco, R.~Gonzalez Felipe, F.~R.~Joaquim and M.~N.~Rebelo,
  {\em Leptogenesis, CP violation and neutrino data: What can we learn?},
  Nucl.\ Phys.\ B {\bf 640} (2002) 202 [hep-ph/0202030];
 
\bibitem{DOnofrio:2014rug}
M.~D'Onofrio, K.~Rummukainen and A.~Tranberg,
{\em Sphaleron Rate in the Minimal Standard Model},
Phys. Rev. Lett. \textbf{113} (2014) no.14, 141602
[arXiv:1404.3565 [hep-ph]].

\bibitem{fy}
M.~Fukugita and T.~Yanagida,
  {\em Baryogenesis Without Grand Unification}, Phys.\ Lett.\ B {\bf 174} (1986) 45.

\bibitem{Blanchet:2012bk}
S.~Blanchet and P.~Di Bari,
{\em The minimal scenario of leptogenesis},
New J. Phys. \textbf{14} (2012), 125012
[arXiv:1211.0512 [hep-ph]].

\bibitem{Casas:2001sr}
J.~A.~Casas and A.~Ibarra,
{\em Oscillating neutrinos and $\mu \to e, \gamma$},
Nucl. Phys. B \textbf{618} (2001), 171-204
[arXiv:hep-ph/0103065 [hep-ph]].
 


\bibitem{Asaka:2005pn}
T.~Asaka and M.~Shaposhnikov,
{\em The $\nu$MSM, dark matter and baryon asymmetry of the universe},
Phys. Lett. B \textbf{620} (2005), 17-26
[arXiv:hep-ph/0505013 [hep-ph]].

\bibitem{Anisimov:2008gg}
A.~Anisimov and P.~Di Bari,
{\em Cold Dark Matter from heavy Right-Handed neutrino mixing},
Phys. Rev. D \textbf{80} (2009), 073017
[arXiv:0812.5085 [hep-ph]].

\bibitem{Dror:2019syi}
J.~A.~Dror, T.~Hiramatsu, K.~Kohri, H.~Murayama and G.~White,
{\em Testing the Seesaw Mechanism and Leptogenesis with Gravitational Waves},
Phys. Rev. Lett. \textbf{124} (2020) no.4, 041804
[arXiv:1908.03227 [hep-ph]].

\bibitem{DiBari:2020bvn}
P.~Di Bari, D.~Marfatia and Y.~L.~Zhou,
{\em Gravitational waves from neutrino mass and dark matter genesis},
Phys. Rev. D \textbf{102} (2020) no.9, 095017
[arXiv:2001.07637 [hep-ph]].

\bibitem{DiBari:2021dri}
P.~Di Bari, D.~Marfatia and Y.~L.~Zhou,
{\em Gravitational waves from first-order phase transitions in Majoron models of neutrino mass},
JHEP \textbf{10} (2021), 193
[arXiv:2106.00025 [hep-ph]].

\bibitem{Fu:2022lrn}
B.~Fu, S.~F.~King, L.~Marsili, S.~Pascoli, J.~Turner and Y.~L.~Zhou,
{\em A predictive and testable unified theory of fermion masses, mixing and leptogenesis},
JHEP \textbf{11} (2022), 072
[arXiv:2209.00021 [hep-ph]].

\bibitem{DiBari:2023mwu}
P.~Di Bari, S.~F.~King and M.~H.~Rahat,
{\em Gravitational waves from phase transitions and cosmic strings in neutrino mass models with multiple majorons},
JHEP \textbf{05} (2024), 068
[arXiv:2306.04680 [hep-ph]].

\bibitem{bounds}
S.~Blanchet and P.~Di Bari,
  {\em New aspects of leptogenesis bounds},
  Nucl.\ Phys.\ B {\bf 807} (2009) 155
  [arXiv:0807.0743 [hep-ph]].

\bibitem{di}
S.~Davidson and A.~Ibarra,
  {\em A Lower bound on the right-handed neutrino mass from leptogenesis},
  Phys.\ Lett.\ B {\bf 535} (2002) 25
  [hep-ph/0202239].

\bibitem{cmb}
W.~Buchmuller, P.~Di Bari and M.~Plumacher,
  {\em Cosmic microwave background, matter - antimatter asymmetry and neutrino masses},
  Nucl.\ Phys.\ B {\bf 643} (2002) 367
   Erratum: [Nucl.\ Phys.\ B {\bf 793} (2008) 362]
  [hep-ph/0205349].
  
\bibitem{Buchmuller:2002jk}
W.~Buchmuller, P.~Di Bari and M.~Plumacher,
{\em A Bound on neutrino masses from baryogenesis},
Phys. Lett. B \textbf{547} (2002), 128-132
doi:10.1016/S0370-2693(02)02758-2
[arXiv:hep-ph/0209301 [hep-ph]].

\bibitem{Buchmuller:2004nz}
W.~Buchmuller, P.~Di Bari and M.~Plumacher,
{\em Leptogenesis for pedestrians},
Annals Phys. \textbf{315} (2005), 305-351
[arXiv:hep-ph/0401240 [hep-ph]].
 
\bibitem{Buchmuller:2003gz}
W.~Buchmuller, P.~Di Bari and M.~Plumacher,
{\em The Neutrino mass window for baryogenesis},
Nucl. Phys. B \textbf{665} (2003), 445-468
[arXiv:hep-ph/0302092 [hep-ph]].

\bibitem{Garbrecht:2024xfs}
B.~Garbrecht and E.~Wang,
{\em The neutrino mass bound from leptogenesis revisited},
JHEP \textbf{03} (2025), 008
[arXiv:2411.09765 [hep-ph]].

\bibitem{plumacher}
M.~Plumacher,
{\em Baryogenesis and lepton number violation},
Z. Phys. C \textbf{74} (1997), 549-559
[arXiv:hep-ph/9604229 [hep-ph]].

\bibitem{geometry} 
P.~Di Bari, {\em Seesaw geometry and leptogenesis},
  Nucl.\ Phys.\ B {\bf 727} (2005) 318
  [hep-ph/0502082].

\bibitem{Barbieri:1999ma}
R.~Barbieri, P.~Creminelli, A.~Strumia and N.~Tetradis,
{\em Baryogenesis through leptogenesis},
Nucl. Phys. B \textbf{575} (2000), 61-77
[arXiv:hep-ph/9911315 [hep-ph]].

\bibitem{flavoureffects}
 A.~Abada, S.~Davidson, F.~-X.~Josse-Michaux, M.~Losada and A.~Riotto,
  {\em Flavor issues in leptogenesis},
  JCAP {\bf 0604} (2006) 004;
E.~Nardi, Y.~Nir, E.~Roulet and J.~Racker,
 {\em The Importance of flavor in leptogenesis},
  JHEP {\bf 0601} (2006) 164.

\bibitem{flavourlep}
S.~Blanchet and P.~Di Bari,
  {\em Flavor effects on leptogenesis predictions},
  JCAP {\bf 0703} (2007) 018
  [hep-ph/0607330].
  
\bibitem{vives}
O.~Vives,
 {\em Flavor dependence of CP asymmetries and thermal 
 leptogenesis with strong right-handed neutrino mass hierarchy},
  Phys.\ Rev.\ D {\bf 73} (2006) 073006
  [hep-ph/0512160].

\bibitem{DiBari:2018fvo}
P.~Di Bari, M.~Re Fiorentin and R.~Samanta,
{\em Representing seesaw neutrino models and their motion in lepton flavour space},
JHEP \textbf{05} (2019), 011
[arXiv:1812.07720 [hep-ph]].

\bibitem{Antusch:2011nz}
S.~Antusch, P.~Di Bari, D.~A.~Jones and S.~F.~King,
{\em Leptogenesis in the Two Right-Handed Neutrino Model Revisited},
Phys. Rev. D \textbf{86} (2012), 023516
[arXiv:1107.6002 [hep-ph]].

\bibitem{Bertuzzo:2010et}
E.~Bertuzzo, P.~Di Bari and L.~Marzola,
{\em The problem of the initial conditions in flavoured leptogenesis and the tauon $N_2$-dominated scenario},
Nucl. Phys. B \textbf{849} (2011), 521-548
[arXiv:1007.1641 [hep-ph]].

\bibitem{DiBari:2014eqa}
P.~Di Bari, S.~King and M.~Re Fiorentin,
{\em Strong thermal leptogenesis and the absolute neutrino mass scale},
JCAP \textbf{03} (2014), 050
[arXiv:1401.6185 [hep-ph]].

 \bibitem{riotto1}
P.~Di Bari and A.~Riotto,
  {\em Successful type I Leptogenesis with SO(10)-inspired mass relations},
  Phys.\ Lett.\ B {\bf 671} (2009) 462
  [arXiv:0809.2285 [hep-ph]].

\bibitem{full}
P.~Di Bari and M.~Re Fiorentin,
  {\em A full analytic solution of $SO(10)$-inspired leptogenesis},
  JHEP {\bf 1710} (2017) 029
  [arXiv:1705.01935 [hep-ph]].

\bibitem{fuller}
S.~Antusch, P.~Di Bari, D.~A.~Jones and S.~F.~King,
 {\em A fuller flavour treatment of $N_2$-dominated leptogenesis},
  Nucl.\ Phys.\ B {\bf 856} (2012) 180
  [arXiv:1003.5132 [hep-ph]].

\bibitem{density}
 S.~Blanchet, P.~Di Bari, D.~A.~Jones and L.~Marzola,
  {\em Leptogenesis with heavy neutrino flavours: from density matrix to Boltzmann equations},
  JCAP {\bf 1301} (2013) 041
  [arXiv:1112.4528 [hep-ph]].

\bibitem{susy}
P.~Di Bari and M.~Re Fiorentin,
 {\em Supersymmetric $SO(10)$-inspired leptogenesis and a new $N_2$-dominated scenario},
  JCAP {\bf 1603} (2016) 039
  [arXiv:1512.06739 [hep-ph]].

\bibitem{ST}
P.~Di Bari and L.~Marzola,
  {\em SO(10)-inspired solution to the problem of the initial conditions in leptogenesis},
  Nucl.\ Phys.\ B {\bf 877} (2013) 719
  [arXiv:1308.1107 [hep-ph]].



\bibitem{shimizu}
Talk by Itaru Shimizu at Neutrino 2024.

\bibitem{preliminary}
Talks by P.~Di Bari and L.~Marzola at the DESY Theory Workshop 2011, 
27-30 September 2011, Hamburg.

\bibitem{DayaBay:2012fng}
F.~P.~An \textit{et al.} [Daya Bay],
{\em Observation of electron-antineutrino disappearance at Daya Bay},
Phys. Rev. Lett. \textbf{108} (2012), 171803
[arXiv:1203.1669 [hep-ex]].

\bibitem{DayaBay:2012fng}
F.~P.~An \textit{et al.} [Daya Bay],
{\em Observation of electron-antineutrino disappearance at Daya Bay},
Phys. Rev. Lett. \textbf{108} (2012), 171803
[arXiv:1203.1669 [hep-ex]].

\bibitem{chianese}
M.~Chianese and P.~Di Bari,
 {\em Strong thermal $SO(10)$-inspired leptogenesis in the light of recent results from long-baseline neutrino experiments}, JHEP {\bf 1805} (2018) 073
  [arXiv:1802.07690 [hep-ph]].

\bibitem{Esteban:2018azc}
I.~Esteban, M.~C.~Gonzalez-Garcia, A.~Hernandez-Cabezudo, M.~Maltoni and T.~Schwetz,
{\em Global analysis of three-flavour neutrino oscillations: synergies and tensions in the determination of $\theta_{23}$, $\delta_{CP}$, and the mass ordering},
JHEP \textbf{01} (2019), 106
[arXiv:1811.05487 [hep-ph]].

\bibitem{Super-Kamiokande:2023ahc}
T.~Wester \textit{et al.} [Super-Kamiokande],
{\em Atmospheric neutrino oscillation analysis with neutron tagging and an expanded fiducial volume in Super-Kamiokande I\textendash{}V},
Phys. Rev. D \textbf{109} (2024) no.7, 072014
[arXiv:2311.05105 [hep-ex]].
 
\bibitem{Ballett:2016daj}
  P.~Ballett, S.~F.~King, S.~Pascoli, N.~W.~Prouse and T.~Wang,
  {\em Sensitivities and synergies of DUNE and T2HK},
  Phys.\ Rev.\ D {\bf 96} (2017) no.3,  033003
  [arXiv:1612.07275 [hep-ph]].

\bibitem{Buchmuller:2001sr}
W.~Buchmuller and M.~Plumacher,
{\em Spectator processes and baryogenesis},
Phys. Lett. B \textbf{511} (2001), 74-76
[arXiv:hep-ph/0104189 [hep-ph]].

\bibitem{Nardi:2005hs}
E.~Nardi, Y.~Nir, J.~Racker and E.~Roulet,
{\em On Higgs and sphaleron effects during the leptogenesis era},
JHEP \textbf{01} (2006), 068
[arXiv:hep-ph/0512052 [hep-ph]].

\bibitem{Babu:2016bmy}
K.~S.~Babu, B.~Bajc and S.~Saad,
{\em Yukawa Sector of Minimal SO(10) Unification},
JHEP \textbf{02} (2017), 136
[arXiv:1612.04329 [hep-ph]].

\bibitem{Babu:2024ahk}
K.~S.~Babu, P.~Di Bari, C.~S.~Fong and S.~Saad,
{\em Leptogenesis in SO(10) with minimal Yukawa sector},
JHEP \textbf{10} (2024), 190
[arXiv:2409.03840 [hep-ph]].

\bibitem{DiBari:2020plh}
P.~Di Bari and R.~Samanta, {\em The $SO(10)$-inspired leptogenesis timely opportunity},
JHEP \textbf{08} (2020), 124
[arXiv:2005.03057 [hep-ph]].

\bibitem{Chen:2025afg}
Z.~Q.~Chen, G.~X.~Fang and Y.~L.~Zhou,
{\em Probing quark-lepton correlation in GUTs with high-precision neutrino measurements},
[arXiv:2511.16196 [hep-ph]].

\bibitem{Fong:2025aya}
C.~S.~Fong and K.~M.~Patel,
{\em Electroweak Triplet Scalar Contribution to $SO(10)$ Leptogenesis},
[arXiv:2505.06391 [hep-ph]].

\end{thebibliography}
\end{document}